\documentclass[aps,prx,onecolumn,superscriptaddress,nobibnotes,nofootinbib]{revtex4-2}

\usepackage{amsthm, color, mathtools, amsmath, amssymb, mathtools, graphicx, float, verbatim, xcolor, placeins, needspace}
\usepackage[caption=false]{subfig}

\usepackage{enumerate}
\usepackage{braket}
\usepackage{url}
\usepackage{algorithm}
\usepackage{algpseudocode}
\usepackage{tikz}
\usetikzlibrary{fit,calc}
\usepackage{microtype} 
\usepackage[british]{babel} 	
\usepackage[unicode=true,bookmarks=true,bookmarksnumbered=false,bookmarksopen=false,breaklinks=false,pdfborder={0 0 1}, backref=false,colorlinks=true]{hyperref}
\addto\captionsbritish{}
\addto\captionsbritish{}
\usepackage{orcidlink}

\newcommand{\todaisci}{Department of Physics, Graduate School of Science, The University of Tokyo, Hongo 7-3-1, Bunkyo-ku, Tokyo 113-0033, Japan}
\newcommand{\todaieng}{Department of Applied Physics, Graduate School of Engineering, The University of Tokyo, 7-3-1 Hongo, Bunkyo-ku, Tokyo, 113-8656, Japan}
\newcommand{\ibmjapan}{IBM Quantum, IBM Research -- Tokyo, Tokyo, Japan}
\newcommand{\ibmus}{IBM Quantum, IBM Thomas J. Watson Research Center, Yorktown Heights, NY 10598}
\newcommand{\manchester}{Department of Physics \& Astronomy, University of Manchester, Manchester M13 9PL, United Kingdom}

\newtheorem{lem}{Lemma}

\newtheorem{Theorem}{Theorem}
\newtheorem{corollary}{Corollary}%[chapter]

\colorlet{pink}{red!40}

\newcommand{\mcE}{\mathcal{E}}
\newcommand{\mcF}{\mathcal{F}}

\newcommand{\mcH}{\mathcal{H}}
\newcommand{\mcI}{\mathcal{I}}

\newcommand{\mcL}{\mathcal{L}}

\newcommand{\mcT}{\mathcal{T}}

\let\oldcomment\algorithmiccomment
\renewcommand{\algorithmiccomment}[1]{\textsf{ \oldcomment{#1}}}

\definecolor{darkgreen}{RGB}{0,200,107}

\definecolor{darkorange}{RGB}{255,100,0}

\begin{document}

\title{Robust Error Accumulation Suppression for Quantum Circuits}

\author{Tatsuki Odake\,\orcidlink{0009-0009-6844-9959}}
\affiliation{\todaisci}

\author{Philip Taranto\,\orcidlink{0000-0002-4247-3901}}
\affiliation{\manchester}
\affiliation{\todaisci}

\author{Nobuyuki Yoshioka\,\orcidlink{0000-0001-6094-8635}}
\affiliation{\todaieng}

\author{\mbox{Toshinari Itoko}\,\orcidlink{0009-0006-2611-2301}} 
\affiliation{\ibmjapan}

\author{Kunal Sharma}
\affiliation{\ibmus}

\author{Antonio Mezzacapo}
\affiliation{\ibmus}

\author{Mio Murao\,\orcidlink{0000-0001-7861-1774}}
\email{murao@phys.s.u-tokyo.ac.jp}
\affiliation{\todaisci}

\date{\today}

\begin{abstract}

We present a \textit{robust error accumulation suppression} (\textbf{REAS}) technique to manage errors in quantum computers. Our method reduces the accumulation of errors in any quantum circuit composed of single- or two-qubit gates expressed as $e^{-i \sigma\theta }$ for Pauli operators $\sigma$ and $\theta \in [0,\pi)$, which forms a universal gate set. For coherent errors---which include gate overrotation and crosstalk---we demonstrate a reduction of the error scaling in an $L$-depth circuit from $O(L)$ to $O(\sqrt{L})$. This asymptotic error suppression behavior can be proven in a regime where all gates---including those constituting the error-suppressing protocol itself---are noisy. Going beyond coherent errors, we derive the general form of decoherence noise that can be suppressed by REAS. Lastly, we experimentally demonstrate the effectiveness of our approach regarding realistic errors using 100-qubit circuits with up to 64 two-qubit gate layers on IBM Quantum processors. 
\end{abstract}

\maketitle

\section{Introduction}
A central problem in quantum information processing is to counteract noise arising from uncontrolled interactions between the principal system and its environment. Such noise causes errors to build up rapidly throughout quantum computations, \emph{a priori} limiting their applicability to logarithmic-depth circuits~\cite{aharonov1996limitations}. Concurrently, the demonstration of quantum algorithms that are exponentially faster than their classical counterparts, e.g., Shor's algorithm~\cite{shor1994algorithms,shor1997algorithm}, has stimulated enormous global efforts towards quantum computing to underpin next-generation technologies~\cite{acin2018quantumtechnologies,awschalom2022roadmap}. 

Developments in quantum error correction provide a promising route to scalable, fault-tolerant quantum devices~\cite{shor1995scheme,calderbank1996good, steane1996error,laflamme1996perfect,terhal2015review,acharya2024google}. Nonetheless, the errors must remain below the ``error correction threshold'' to provide any practical benefit. For moderate logical gate error rates (on the order of $10^{-8}$), the errors accumulate at a rate that is insurmountable by correction procedures alone; thus, one must incorporate other methods of dealing with errors beyond correction.

To this end, two classes of complementary approaches are widely considered, namely \textit{error mitigation} and \textit{error suppression}. Quantum error mitigation aims to reduce the bias in estimating expectation values of physical observables at the cost of increased variance~\cite{temme2017error, li2017efficient, endo2018practical, koczor_exponential_2021, huggins_virtual_2021, yoshioka2022generalized, hakoshima2023localized,cai2023review}. Incorporating such methods has recently led to quantum simulations that are competitive with state-of-the-art classical ones~\cite{kandala2017hardware, kandala2019error,kim2023scaleable,kim2023evidence}. Nevertheless, a crucial drawback of error mitigation is the exponential increase in the number of measurements required~\cite{tsubouchi2023universal, takagi2023universal,quek2022exponentially}. 

Quantum error suppression takes a different approach: instead of shifting the burden to measurement overheads, here one modifies the noisy circuit to design a quantum operation that approximates the target one. A paradigmatic example is the well-known dynamical decoupling protocol, which rectifies decoherence effects on idle qubits by periodically repeating fast random pulses that delete any spuriously imprinted information~\cite{viola1999dynamical}. However, dynamical decoupling only serves to keep the state of the system from decohering, and it is a priori unclear how to extend it to situations where noisy gates are being applied to a physical system for the purpose of a computation. 

Fortunately, in such scenarios, coherent errors in circuits composed of Clifford gates---e.g., accidental gate overrotation---can be suppressed via related techniques that beneficially harness randomness, e.g., probabilistic compiling~\cite{hastings2017turning, campbell_shorter_2017, akibue2024probabilistic} and Pauli twirling~\cite{dur2005standard, magesan2011scalable,geller2013efficient}. More importantly, the procedure of randomized compiling applies beyond such limitations and can even suppress the growth of errors in universal (i.e., non-Clifford) circuits~\cite{wallman2016noise, hines2023demonstrating, santos2024deeper}. Nonetheless, the class of noise that can be counteracted via such methods remains somewhat restricted and may be surpassed in realistic near-term settings, such as partially fault-tolerant architectures with analog rotation gates~\cite{akahoshi2023partially}. Secondly, it is not clear that known techniques are robust against errors in their own application (as opposed to those occurring in the computation gates). Lastly, the extent to which error suppression techniques can counteract decoherence gate-level errors (while keeping the desired gate applied) is yet to be fully determined in general, although experimental evidence suggests that such errors can be somewhat dealt with~\cite{hashim2021randomized,ville2022leveraging,ferracin2022efficiently,kurita2023synergetic,goss2023extending,kim2023scaleable,kim2023evidence}. 

Here, to overcome these problems, we present an advanced procedure for managing errors in quantum computers: \textit{robust error accumulation suppression} (\textbf{REAS}). Our method reduces the accumulation of errors in any circuit composed of single- or two-qubit gates expressed as $e^{-i \sigma\theta }$ for Pauli operators $\sigma$ and $\theta \in [0,\pi)$, which forms a universal gate set. Improving upon standard error suppression methods, our technique reduces the accumulation of first-order error terms whilst keeping the variance unchanged, leveraging a nested and correlated structure of the inserted Pauli gates. In the case of coherent errors---which includes spectator crosstalk~\cite{sundaresan2020reducing}---we demonstrate a reduction of the error scaling in an $L$-depth circuit from $O(L)$ to $O(\sqrt{L})$. Crucially, REAS makes no assumption on the cleanness of the error-suppressing protocol itself and is, therefore, truly robust, applying to situations in which the newly inserted gates themselves are noisy. Furthermore, we detail the class of decoherence noise that can be suppressed via REAS, namely that which is well-approximated by weak system-environment coupling, and demonstrate numerical evidence for the quadratic suppression of error scaling. 

This paper is outlined as follows. In Sec.~\ref{sec::reas}, we formalize our framework and analyze a motivating example. In Sec.~\ref{sec::results}, we present our method for robustly suppressing the accumulation of errors and prove our main theoretical results. To highlight the applicability and effectiveness of our approach, we supplement this analysis with numerical simulations in Sec.~\ref{sec::numericalresults}, as well as provide an experimental demonstration using IBM Quantum processors in Sec.~\ref{sec::experiments}.

%%%%%%%%%%%%%%%%%%%%%%%%%%%%%%%%%%%%%%%%%%%%%%

%%%%%%%%%%%%%%%%%%%%%%%%%%%%%%%%%%%%%%%%%%%%%%

\section{Robust Error Accumulation Suppression}\label{sec::reas}

\subsection{Layered Circuits}\label{subsec::layeredcircuits}

\begin{figure}[t]
    \centering
\includegraphics[width=0.45\linewidth]{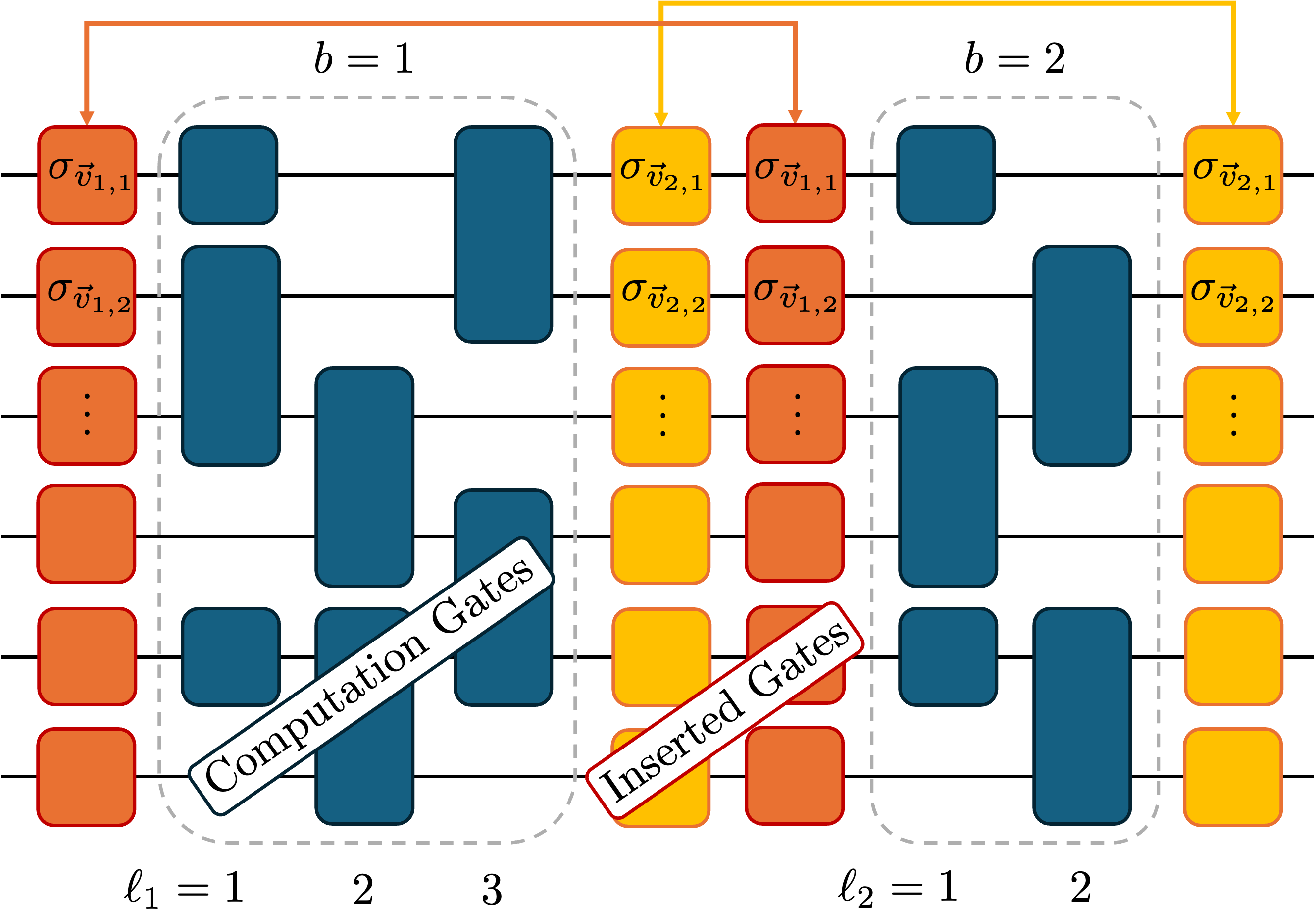}
    \caption{\textit{REAS Schematic.---}An $n=6$ qubit quantum circuit, consisting of $B=2$ blocks (gray dashed outline) of computation gates (blue), with the first block comprising three layers and the second comprising two layers. Each such computation gate acts on at most two qubits and is of the form $e^{-i\sigma\theta}$ for $\theta \in [0,\pi)$. Between each block $b$, a random single qubit Pauli gate $e^{-i \sigma_{\vec{v}_{b, q}}}$ is applied to qubit $q$ (orange, yellow). The nested, correlated structure of these inserted gates (depicted by the arrows at the top) leads to a quadratic suppression of the accumulation of errors (in terms of circuit length). This error suppression works even in the case where the inserted gates themselves are noisy, since every gate (including the inserted random Pauli gates) is itself sandwiched by correlated random Pauli gates (e.g., the leftmost yellow Pauli gates are sandwiched by the orange Pauli gates).\vspace{-1em} }
    \label{fig::reas_schematic}
\end{figure}

We begin by presenting the framework that describes circuit-based quantum computation with errors. Suppose that one wishes to implement a layered circuit consisting of single- or two-qubit gates of the form $e^{-i\sigma\theta}$ for rotation angle $\theta \in [0,\pi)$ (which includes all gates of the form $e^{-i\sigma\theta' }$ for $\theta' \in \mathbb{R}$ up to a global phase). Consider such a circuit of $L$ layers of gates applied to an $n$-qubit system; in each independent layer $\ell$, the particular qubits that are acted on non-trivially are denoted by $Q_\ell$, which consists of disjoint subsets of $\{ 1, \hdots, n\}$. Grouping the layers into $B$ blocks (which can now comprise sequentially implemented gates on certain qubits), leads to
\begin{align}\label{eq::desc_of_layered_block}
        \prod_{b=1}^{B} \prod_{\ell_b} \left(\bigotimes_{\vec{q} \in Q_{\ell_b}} e^{-i\sigma_{\ell_b,\vec{q}}\,\theta_{\ell_b,\vec{q}}}
        \right),
\end{align}
where $\ell_b$ labels the layer \textit{within} each block $b$. For a block with $L_B$ layers, the total number of layers is $BL_B = L$; throughout, we consider $B \propto L$, so the scaling behavior in terms of layers or blocks is identical up to constant factors. We refer to the Pauli operators $\sigma_{\ell_b, \vec{q}}$ inside the exponential in the circuit as ``computation gates'' to distinguish them from other types of Pauli gates that will be inserted at will in order to suppress potential errors in the computation, which we will term ``inserted gates'' (see Fig.~\ref{fig::reas_schematic}).

\subsection{Motivating Example}\label{subsec::motivatingexample}

To motivate the problem at hand, we now analyze the impact of (small) coherent gate errors on block-layered quantum circuits of the form in Eq.~\eqref{eq::desc_of_layered_block}. Intuitively, the actual operation applied to each layer is shifted slightly with respect to the desired one by some additional unitary rotation, leading to the implementation of
\begin{align}
\left(\bigotimes_{\vec{q} \in Q_{\ell_b}} e^{-i\sigma_{\ell_b,\vec{q}}\,\theta_{\ell_b,\vec{q}}}
        \right)\left[ I+i\epsilon C_{\ell_b} \right],
\end{align}
where $\| C_{\ell_b} \|_{\rm op}\leq 1\ $ for all $b$ and $\ell_b$, and $\epsilon>0$ is small such that $I+i\epsilon C_{\ell_b}$ is close to the identity, encoding the assumption that the error in each layer is not too large. In any such layered circuit, there are $L = \sum_{b} \ell_b$ layers where $O(\epsilon)$ errors can occur, and thus the total error can be $O(L\epsilon)$ in the worst case.

Inspired by the method of Pauli twirling, we now consider the effect of inserting random Pauli operators on all qubits between each block. Let $\sigma_{\vec{v}_b} := \bigotimes_{k=1}^{n} \sigma_{v_b}^{(k)}$ for $v_b\in \{0,1,2,3\}$ be a uniformly randomly sampled $n$-qubit Pauli operator applied before block $b$. Consider an arbitrary block $b$ preceded by some string of inserted Pauli gates $\sigma_{\vec{v}_b}$. Now, for tensor products of (inserted) qubit Pauli gates, any such overall operation either commutes or anti-commutes with any (one or two-qubit) Pauli gates $e^{-i\sigma_{\ell_b, \vec{q}} \theta_{\ell_b, \vec{q}}}$ applied in the subsequent block. This permits us to define
\begin{align}
    s:=
    \begin{cases}
        1& \quad\quad \text{if } \sigma_{\ell_b, \vec{q}}\text{ and }\sigma_{\vec{v}_b}\text{ commute}\\
        -1& \quad \quad \text{if } \sigma_{\ell_b, \vec{q}}\text{ and }\sigma_{\vec{v}_b}\text{ anti-commute.}
    \end{cases}
\end{align}
Although we do not explicitly label it for ease of notation, $s$ inherently depends upon $\ell_b, \vec{q}$ and $\vec{v}_b$: each computational gate $\sigma_{\ell_b, \vec{q}}$ in the bock may have a different sign factor $s$ depending on its commutation relation with the specific inserted Pauli string $\sigma_{\vec{v}_b}$ (so strictly speaking, $s = s_{\ell_b,\vec{q},\vec{v}_b}$). 

With this, we can express the circuit transformation between blocks succinctly as follows. When a random Pauli string $\sigma_{\vec{v}_b}$ is inserted before block $b$, the computational gates within the block are sandwiched by said Pauli string:
\begin{align}\label{eq::changing_rot_direction}
    \prod_{\ell_b} e^{-i\sigma_{\ell_b, \vec{q}}\theta_{\ell_b, \vec{q}}} \mapsto \sigma_{\vec{v}_b} \left( \prod_{\ell_b} e^{-is\sigma_{\ell_b, \vec{q}}\theta_{\ell_b, \vec{q}}} \right) \sigma_{\vec{v}_b}.
\end{align}
Here, the arrow $\mapsto$ indicates that we rewrite the original computational circuit into this equivalent form by appropriately conjugating with the inserted Pauli gates. The factor $s = \pm 1$ accounts for whether the inserted Pauli string $\sigma_{\vec{v}_b}$ commutes ($+1$) or anti-commutes ($-1$) with its neighboring computational Pauli gates $\sigma_{\ell_b, \vec{q}}$. Note that the technique of transforming computational Pauli gates to the form of Eq.~\eqref{eq::changing_rot_direction} itself is not new; however, in previous works, the error on the inserted Pauli gate $\sigma_{\vec{v}_b}$, as well as the possible dependence of such errors on the rotation direction through $s$, was not explicitly considered. Here, we demonstrate that such errors can also be dealt with, highlighting that our method is robust against coherent noise \emph{in its own implementation}.

Consider now for the sake of simplicity single-layer blocks, in which case we simply write $\sigma_{\ell_b, \vec{q}} =: \sigma$ and $\theta_{\ell_b, \vec{q}} =: \theta$ for clarity. In the absence of any errors, both the original gate $e^{-i\sigma\theta}$ and the average transformed one $(1/4)^n\sum_{\vec{v}_b}\sigma_{\vec{v}_b}e^{-is\sigma\theta}\sigma_{\vec{v}_b}$ lead to the same overall gate applied. However, whenever the computation gates $e^{i s \sigma\theta}$ are affected by a coherent error (i.e., corresponding to a unitary rotation of the form $I+i\epsilon C_s$), the original gate suffers from an $O(\epsilon)$ error, whereas the transformed gate undergoes the change
\begin{align}\label{eq::motivatingcoherenterror}
    \left(\frac{1}{4}\right)^n
    \sum_{\vec{v}_b}\sigma_{\vec{v}_b}
    \left[
    e^{-is\sigma\theta}(I+i\epsilon C_s)
    \right]
    \sigma_{\vec{v}_b}
    &=\left(\frac{1}{4}\right)^n
    \sum_{\vec{v}_b}
    e^{-i\sigma\theta}[ I+i\epsilon 
    \sigma_{\vec{v}_b}C_s\sigma_{\vec{v}_b}
    ] \notag \\
    &=e^{-i\sigma\theta}[I+i\epsilon(\alpha I+ \beta \sigma)]
    =e^{i\epsilon \alpha}e^{-i(\theta-\epsilon \beta)\sigma}+O(\epsilon^2),
\end{align}
where $\alpha$ and $\beta$ are the coefficients of the term in $C_s$ proportional to $I$ and $s\sigma$, respectively. This can be shown from
\begin{align}\label{eq::main-randompauliaverageidentity}
    \left(\frac{1}{4^n}
    \right)
    \sum_{\vec{v}_b\in\{0,1,2,3\}^n}
    \sigma_{\vec{v}_b}
    \sigma
    \sigma_{\vec{v}_b}
    =\begin{cases}
        I& \quad\quad \text{for }\sigma=I\\
        0& \quad\quad \text{for }\sigma\neq I
    \end{cases}
\end{align}
and 
\begin{align}\label{eq::main-no_para_s}
    \left(\frac{1}{4^n}
    \right)
    \sum_{\vec{v}_b\in\{0,1,2,3\}^n}
    \sigma_{\vec{v}_b}
    s\sigma'
    \sigma_{\vec{v}_b}=
    \begin{cases}
        \sigma& \quad\quad \text{for }\sigma=\sigma'\\
        0& \quad\quad \text{for }\sigma\neq\sigma'.
    \end{cases}
\end{align}
Since $e^{i\epsilon \alpha}$ in Eq.~\eqref{eq::motivatingcoherenterror} only contributes to a global phase and can thus be ignored, the additional $O(\epsilon)$ shift $-\epsilon \beta$ in the rotation angle is the only error that must be dealt with in order to erase all $O(\epsilon)$ error within each layer. 

The general (multi-layer block) expression for how the overall layered circuit with coherent errors transforms on average upon inserting random Pauli gates between blocks is 
\begin{align}\label{eq::desc_of_layered-2}
    \left(\frac{1}{4^n}
    \right)^B \sum_{\vec{v}_1,\ldots ,\vec{v}_B}
    \left[\sigma_{\vec{v}_B}  \prod_{\ell_B} \left(\bigotimes_{\vec{q} \in Q_{\ell_B}} e^{-is \, \sigma_{\ell_B,\vec{q}} \,\theta_{\ell_B,\vec{q}}}
        \right) \right]
    \prod_{b=1}^{B-1} \left[\sigma_{\vec{v}_{b+1:b}}  \prod_{\ell_b} \left(\bigotimes_{\vec{q} \in Q_{\ell_b}} e^{-is\,\sigma_{\ell_b,\vec{q}}\,\theta_{\ell_b,\vec{q}}}
        \right) \right] \sigma_{\vec{v}_1} \notag \\
        = \prod_{b=1}^{B} \prod_{\ell_b} \left(\bigotimes_{\vec{q} \in Q_{\ell_b}} e^{-i\sigma_{\ell_b,\vec{q}}\,(\theta_{\ell_b,\vec{q}}+ \epsilon \beta_{\ell_b,\vec{q}})}
        \right) + O(\epsilon + B \epsilon^2).
\end{align}
Here, we have introduced the notation $\vec{v}_{j:k}$ for two vectors that satisfy $\sigma_{\vec{v}_{j:k}} \propto \sigma_{\vec{v}_k}\sigma_{\vec{v}_j}$ and implicitly understand the value of $s$ to depend upon whether or not each inserted Pauli gate string commutes or anti-commutes with each layer of computational gates in the neighboring block. We refer to the random Pauli operators $\sigma_{\vec{v}_1},\sigma_{\vec{v}_{2:1}}\ldots,\sigma_{\vec{v}_B}$ as the ``inserted gates'' to distinguish them from the computational Pauli gates. Note the {\it nested and correlated} structure of the inserted random Pauli gates: each inserted gate $\sigma_{\vec{v}_j}$ is sandwiched by either $\sigma_{\vec{v}_{j-1}}$ or $\sigma_{\vec{v}_{j+1}}$ for $1<j<B$, which is what provides our scheme with robustness against noise in the inserted gates themselves.

In the following, we will demonstrate a method that cancels the $O(\epsilon)$ errors caused by the shift in rotation angle. This leaves an overall error that scales as $O(\epsilon + B\epsilon^2)$, which approaches $O(B\epsilon^2)$ for circuits with high gate depth. Importantly, our procedure also works when the inserted operations themselves are noisy (which we have not explicitly included above but will below), thereby demonstrating robustness of the method. \vspace{-1em}

%%%%%%%%%%%%%%%%%%%%%%%%%%%%%%%%%%%%%%%%%%%%%%%%%

\section{Main Results}\label{sec::results}

\subsection{Robust Error Accumulation Suppression}\label{subsec::reas}

We now present our method of \emph{robust error accumulation suppression} (\textbf{REAS}). This procedure comprises of first performing a \emph{calibration} process by estimating the rotation error in each layer, and then \emph{inserting random Pauli gates} in order to counteract this in a tailored manner, removing all $O(\epsilon)$ errors as we will show in Sec.~\ref{subsec::coherent}. An optional third step involves transforming all single-qubit computational gates into two-qubit ones in such a way that certain types of decoherence errors can also be suppressed, as we will discuss in Sec.~\ref{subsec::decoherence}. Details are provided in App.~\ref{app::reascalibration}. 

\textbf{Calibration.} The calibration step is performed by considering each layer of the circuit individually. For each chosen layer $\ell_b$, one repeats the computational gates of that layer $K$ times and intersperses it with random Pauli gate strings, which leads to the transformation\footnote{Note that the value of $s$ in each term here differs, as it depends explicitly upon the commutation relation with the Pauli gates inserted at its neighboring position.} 
\begin{align}\label{eq::reas_calibration}
    \left(\frac{1}{4^n}
    \right)^K \sum_{\vec{v}_1,\ldots ,\vec{v}_K}
    \sigma_{\vec{v}_K} \left( \bigotimes_{\vec{q} \in Q_{\ell_b}} e^{-is \, \sigma_{\ell_b,\vec{q}} \,\theta_{\ell_b,\vec{q}}}
        \right) 
    \prod_{k=1}^{K} \left[\sigma_{\vec{v}_{k+1:k}} \left(\bigotimes_{\vec{q} \in Q_{\ell_b}} e^{-is\,\sigma_{\ell_b,\vec{q}}\,\theta_{\ell_b,\vec{q}}}
        \right) \right] \sigma_{\vec{v}_1}.
\end{align}
Since the undesired shift in the rotation angle accumulates by repeating the same layer $K$ times, the value of $\epsilon \beta_{\ell_b,\vec{q}}$ is a systematic (i.e., additive) error and can therefore be estimated via robust phase estimation~\cite{kimmel2015robust}. 

This technique uses a robust variant of the well-known phase estimation procedure~\cite{Higgins_2009} for determining such systematic errors in a universal single-qubit gate set to achieve optimal efficiency, i.e., Heisenberg scaling, without requiring entanglement or auxilliary systems. Phase estimation is a metrological technique that works by performing sequences of measurements to obtain observables from which systematic errors such as gate overrotation can be extracted, thereby providing a foundation for gate calibration. The power of this method comes from the fact that small errors in gates coherently accumulate into large observables when applied in sequence many times, allowing them to be efficiently estimated. In the standard version, state preparations and measurements must be performed perfectly, and gates must be of an assumed form. However, the robust variant relaxes these conditions and permits imperfections; nonetheless, Heisenberg scaling is still achievable (see App.~\ref{app::robustphaseestimation}). 

In order to obtain an estimate with at most $O(\epsilon^2)$ error, $K$ should be taken to be $O(1/\epsilon^2)$. Note finally that the error term of $O(\epsilon +B\epsilon^2)$ in Eq.~\eqref{eq::desc_of_layered-2} can still be small enough to effectively perform robust phase estimation by taking $B$ proportional to $1/\epsilon^2$. 

\textbf{Inserting Random Pauli Gates.} Using the value of the rotation error shifts obtained during the calibration procedure, one can then insert random Pauli gates into the calibrated layered circuit to specifically suppress the errors in each block (see Fig.~\ref{fig::reas_schematic}). This is achieved by shifting the rotation angle of all computational Pauli gates in the original circuit by angles of the same magnitude as that estimated in the calibration but of the opposite sign to cancel out the error.

It is straightforward to see that our method works to suppress errors in non-Clifford circuits since the values of the rotation angles above are not necessarily from a discrete set. Moreover, as we will demonstrate in the coming section, additional coherent errors on the inserted Pauli gates can also be dealt with, thereby endowing our method with robustness that is crucial in realistic settings.\vspace{-1em}

\subsection{Suppressing the Accumulation of Coherent Errors}\label{subsec::coherent}

The REAS scheme suppresses any coherent error, both on the computational Pauli gates and on the inserted ones such that the error on the overall circuit $\prod_{b=1}^B \prod_{\ell_b} \left(\bigotimes_{\vec{q} \in Q_{\ell_b}} e^{-i\sigma_{\ell_b,\vec{q}}\theta_{\ell_b,\vec{q}}}\right)$ decreases from $O(\epsilon B)$ to $O(\epsilon+\epsilon^2B)$ (assuming that the maximum size of any block is upper bounded by a constant, as we will do throughout).

Up to now, we have analyzed REAS in the setting where one averages over the inserted random Pauli gates. Although it might seem at first glance that the average case being close to the ideal one is not necessarily related to the error on any individual gate, in App.~\ref{app::diamondnorm}, we show that the former condition implies that the \textit{root mean square} (\textbf{RMS}) of the latter is small, thereby bounding ``worst case'' scenarios. In particular, we show that the RMS error (using the diamond distance) between the error-free unitary implemented by $\prod_{b=1}^B \prod_{\ell_b} \left(\bigotimes_{\vec{q} \in Q_{\ell_b}} e^{-i\sigma_{\ell_b,\vec{q}}\theta_{\ell_b,\vec{q}}}\right)$ and any single-shot unitary operator implemented after calibration in a given run with the random Pauli gate string inserted scales as $O(\sqrt{\epsilon+\epsilon^2B})$, which approaches $O(\epsilon \sqrt{B})$ in the regime $B\gg 1/\epsilon$. 

Now, we briefly explain how the REAS suppresses any coherent error occurring on the implementation of the inserted random Pauli operators themselves. Suppose that a single error term $i\epsilon C_{\vec{v}_{b'+1;b'}}$ occurs on a Pauli operator $\sigma_{\vec{v}_{b'+1;b'}}$ of 
\begin{align}
    \left(\frac{1}{4^n}
    \right)^B \sum_{\vec{v}_1,\ldots ,\vec{v}_B}
    \left[\sigma_{\vec{v}_B}  \prod_{\ell_B} \left(\bigotimes_{\vec{q} \in Q_{\ell_B}} e^{-is \, \sigma_{\ell_B,\vec{q}} \,\theta_{\ell_B,\vec{q}}}
        \right) \right]
    \prod_{b=1}^{B-1} \left[\sigma_{\vec{v}_{b+1:b}}  \prod_{\ell_b} \left(\bigotimes_{\vec{q} \in Q_{\ell_b}} e^{-is\,\sigma_{\ell_b,\vec{q}}\,\theta_{\ell_b,\vec{q}}}
        \right) \right] \sigma_{\vec{v}_1},
\end{align}
so that the overall gate sequence implemented is
\begin{align}
    &\left(\frac{1}{4^n}
    \right)^B \sum_{\vec{v}_1,\ldots ,\vec{v}_B}
    \nonumber\\
    &\sigma_{\vec{v}_B}\left[\cdots \right]
    \sigma_{\vec{v}_{b'+2;b'+1}}
    \prod_{\ell_{b'+1}}
    \left(
    \bigotimes_{\vec{q}\in Q_{\ell_{b'+1}}}
    e^{-is\sigma_{\ell_{b'+1},\vec{q}}\theta_{\ell_{b'+1},\vec{q}}}
    \right)
    (\sigma_{\vec{v}_{b'+1;b'}}i\epsilon C_{\vec{v}_{b'+1;b'}})
    \prod_{\ell_{b'}}
    \left(
    \bigotimes_{\vec{q}\in Q_{\ell_{b'}}}
    e^{-is\sigma_{\ell_{b'},\vec{q}}\theta_{\ell_{b'},\vec{q}}}
    \right)
    \sigma_{\vec{v}_{b';b'-1}}
    \left[\cdots\right]\sigma_{\vec{v}_1}.
\end{align}
This expression can be rewritten as
\begin{align}
    &\left(\frac{1}{4^n}
    \right)^2\sum_{\vec{v}_{b'},\vec{v}_{b'+1}}
    \prod_{b=b'+1}^{B}
    \prod_{\ell_b} \left(\bigotimes_{\vec{q} \in Q_{\ell}} e^{-i\sigma_{\ell_b,\vec{q}}\theta_{\ell_b,\vec{q}}}\right)
    \sigma_{\vec{v}_{b'}}i\epsilon C_{\vec{v}_{b':b'+1}}\sigma_{\vec{v}_{b'}}
    \prod_{b=1}^{b'}
    \prod_{\ell_b} \left(\bigotimes_{\vec{q} \in Q_{\ell}} e^{-i\sigma_{\ell_b,\vec{q}}\theta_{\ell_b,\vec{q}}}\right)
    \nonumber\\
    \quad\quad&=[\cdots]
    \left(\frac{1}{4^n}
    \right)
    \sum_{\vec{v}_{b'}}\sigma_{\vec{v}_{b'}}
    \left(\frac{1}{4^n}
    \sum_{\vec{v}_{b'+1}}i\epsilon C_{\vec{v}_{b':b'+1}}\right)\sigma_{\vec{v}_{b'}}
    [\cdots] =O(\epsilon^2),
\end{align}
where $[\cdots]$ represents the terms that are irrelevant. As shown in App.~\ref{app::reascalibration}, this overall expression is $O(\epsilon^2)$ since $\sum_{\vec{v}_{b'+1}}C_{\vec{v}_{b':b'+1}}$ is independent of $\vec{v}_{b'}$ and the term in $i\epsilon C_{\vec{v}_{b':b'+1}}$ proportional to the identity can be taken to be $O(\epsilon^2)$ without loss of generality. 

Lastly, in App.~\ref{app::reasvsrc}, we analyze the relation between REAS and \emph{randomized compiling} (\textbf{RC})~\cite{wallman2016noise}. Although both methods make use of inserting random gates to suppress errors, they do so in different ways, leading to distinct scaling behaviors. In REAS, the inserted gates depend upon their commutation relation with their neighboring gates, in contrast to those in RC which automatically cancel coherent errors by averaging over Pauli operators. Thus, the REAS approach requires additional necessity to deal with $s_k$-dependent errors on gates of the form $e^{\pm i \sigma_k \theta_k}$; the aforementioned calibration and single Pauli transformation subroutines can deal with such errors (but are based upon a particular model of decoherence error; see below and App.~\ref{app::decoherence}). Of course, such dependence, together with the fact that REAS can increase the depth of the circuit (albeit only to at most twice as long), can be thought of as a disadvantage of REAS compared to RC. On the other hand, these differences makes REAS more robust against gate-dependent errors on the twirling gates. 

\subsection{Suppressing the Accumulation of Decoherence Errors (with Single Pauli Transformation)}\label{subsec::decoherence}

The original procedure of REAS comprising calibration and then inserting random Pauli gates cannot immediately suppress general error due to decoherence. However, we now incorporate an additional step, which we dub ``single Pauli transformation'', that allows our method to deal with certain classes of decoherence errors, in particular those modeled by weak system-environment interactions. See App.~\ref{app::decoherence} for details.

\textbf{Single Pauli Transformation.} Here we modify all single-qubit computation gates $e^{-is\sigma_{l_b,\vec{q}}\theta_{l_b,\vec{q}}}$ (i.e., those except for the inserted random Pauli operators) as follows
\begin{align}\label{eq::singlepaulitransformation}
&e^{isYIt}\to e^{i\tfrac{\pi}{4}XZ}e^{isZZt}e^{-i\tfrac{\pi}{4}XZ}
\nonumber\\
&e^{isZIt}\to e^{-i\tfrac{\pi}{4}XI}e^{i\tfrac{\pi}{4}XZ}e^{isZZt}e^{-i\tfrac{\pi}{4}XZ}e^{i\tfrac{\pi}{4}XI}
\nonumber\\
&e^{isXIt}\to e^{i\tfrac{\pi}{4}ZI}e^{i\tfrac{\pi}{4}XZ}e^{isZZt}e^{-i\tfrac{\pi}{4}XZ}e^{-i\tfrac{\pi}{4}ZI}.
\end{align}
This transformation makes the single-qubit Pauli rotation gate robust against decoherence errors, thereby enabling the tolerance of the REAS procedure against such noise. The trading of such decoherence errors for gate errors comes at the cost of introducing longer sequences of two-qubit gates; however, since these sequences increased by a constant factor, in certain situations, the scaling behavior will become favorable. As we will show below, this mapping leads to the suppression of decoherence errors that are up to first order in the operator norm of the interaction Hamiltonian.

Upon invoking the single Pauli transformation, the error-suppressing properties of REAS against decoherence errors become similar to those explained previously. In particular, this procedure suppresses decoherence error on both the computational gates and the inserted random Pauli operators in such a way that the error on the overall circuit $\prod_{\ell_b=1}^{B} \left(\bigotimes_{\vec{q} \in Q_{\ell_b}} e^{-i\sigma_{\ell_b,\vec{q}}\theta_{\ell_b,\vec{q}}}\right)$ changes from $O(\epsilon B)$ to $O(\epsilon+\epsilon ^2 B)$ in the average case; similarly, the RMS of the diamond distance between the error-free unitary operation implemented by $\prod_{\ell_b=1}^B \left(\bigotimes_{\vec{q} \in Q_{\ell_b}} e^{-i\sigma_{\ell_b,\vec{q}}\theta_{\ell_b,\vec{q}}}\right)$ and any single-shot unitary operator implemented after calibration in a given run with random Pauli gates inserted scales as $O(\sqrt{\epsilon+\epsilon^2B})$.

We now move on to analyze the class of decoherence errors that can be dealt with via REAS. We model the decoherence error via introducing an interaction Hamiltonian $H'_{\ell_b,\vec{q},s_{\vec{q}}}(t)\in \mcH\otimes \mcH_{\rm env}$ of order $\epsilon$ that leads to the pulse Hamiltonian that implements the desired $e^{-is_{\vec{q}}\sigma_{\ell_b,\vec{q}}\theta_{\ell_b,\vec{q}}}$. Note that, since $H'_{\ell_b,\vec{q},s_{\vec{q}}}(t)$ can contain arbitrary terms that lead to both coherent and incoherent errors, our statements here are a direct generalization of those made in the previous section regarding coherent errors. Note, however, that this approach requires engineered interactions with auxiliary qubits in order to suppress the decoherence error; in other words, the decoherence process must be engineered or at least accurately modeled rather than stemming from truly spurious noise.

We begin by assuming that the decoherence noise applied to the layer $\left(\bigotimes_{\vec{q} \in Q_{\ell_b}} e^{-is_{\vec{q}}\sigma_{\ell_b,\vec{q}}\theta_{\ell_b,\vec{q}}}\right)$ can be expressed as
\begin{align}
    \mcT\exp\left[
    -i\int_0^T {\rm d}t' \left(\sum_{\vec{q}\in Q_{\ell_b}}H_{\ell_b, \vec{q},s_{\vec{q}}}(t')\right)+ H_{{\rm env}}+ \sum_{\vec{q}\in Q_{\ell_b}}H'_{\ell_b, \vec{q},s_{\vec{q}}}(t')
    \right],
\end{align}
where $H_{\ell_b, \vec{q},s_{\vec{q}}}(t') \in \mcH$ is the time-dependent pulse Hamiltonian to implement the desired gate $e^{-is_{\vec{q}}\sigma_{\ell_b, \vec{q}}\theta_{\ell_b,\vec{q}}}$, $H'_{\ell_b, \vec{q},s_{\vec{q}}}(t') \in \mcH\otimes\mcH_{\rm env}$ is the undesired Hamiltonian that is induced when applying $H_{\ell_b, \vec{q},s_{\vec{q}}}$ in the presence of noise, $H_{\rm env} \in \mcH_{\rm env}$ is the background Hamiltonian of the environment, and $\mathcal{T}$ is the time ordering operator. Again, identity operators on subspaces where the dynamics acts trivially are implied. We also assume that the undesired Hamiltonian $H'_{\ell_b, \vec{q},s_{\vec{q}}}(t')$ does not contain interaction terms involving more than two systems (either qubits in $\mcH$ or systems in $\mcH_{\rm env}$); since most Hamiltonians in nature involve at most two-body interactions, this assumption is natural.

Note that many relevant decoherence mechanisms---such as depolarizing, amplitude damping, and dephasing---can be expressed by this interaction model. Indeed, for single-qubit decoherence channels
\begin{align}
\mathcal{E}^{\mathrm{dpl}}_p(\rho ):=
(1-p)\rho +p\frac{I}{2}, \quad \quad \quad
\mathcal{E}^{\mathrm{ad}}_{\gamma }(\rho ):=
E_0\rho E_0^{\dagger}+E_1 \rho E_1^{\dagger}, \quad \quad \quad 
\mathcal{E}^{\mathrm{dph}}_p(\rho ):=(1-p)\rho +pZ\rho Z,
\end{align}
where
\begin{align}
    &E_0:=\left(\begin{array}{cc}
1&0\\
0&\sqrt{\gamma}\\
\end{array} \right) \quad \text{and} \quad
E_1:=\left(\begin{array}{cc}
0&\sqrt{\gamma }\\
0&0\\
\end{array} \right) ,
\end{align}
their action can be modeled explicitly as
\begin{align}
    \mathcal{E}^{\mathrm{dpl}}_{\sin^2 (2\delta t)}(\rho)
    &=
    {\rm tr}_{\rm env}\left[e^{-i\delta t(X\otimes X_{\mathrm{env}}+Y\otimes Y_{\mathrm{env}}+Z\otimes Z_{\mathrm{env}})}\left(\rho\otimes \frac{I_{\rm env}}{2}\right)
    e^{i\delta t(X\otimes X_{\mathrm{env}}+Y\otimes Y_{\mathrm{env}}+Z\otimes Z_{\mathrm{env}})}
    \right]
    \nonumber\\
\mathcal{E}^{\mathrm{ad}}_{\sin^2 (\delta t)}(\rho)&=
    {\rm tr}_{\rm env}\left[e^{-i\tfrac{\delta t}{2}(X\otimes X_{\mathrm{env}}+Y\otimes Y_{\mathrm{env}})}(\rho\otimes \ket{0}\bra{0}_{\rm env})e^{i\tfrac{\delta t}{2}(X\otimes X_{\mathrm{env}}+Y\otimes Y_{\mathrm{env}})}\right]
    \nonumber\\
    \mathcal{E}^{\mathrm{dph}}_{\sin^2 (\delta t)}(\rho)&=
    {\rm tr}_{\rm env}\left[e^{-i\delta t(Z\otimes Z_{\mathrm{env}})}\left(\rho\otimes \frac{I_{\rm env}}{2}\right)
    e^{i\delta t(Z\otimes Z_{\mathrm{env}})}
    \right] .
\end{align}
As is the case regarding dynamical decoupling, note that different ways of modeling decoherence in this interaction picture can lead to varying performance of REAS. For instance, in a similar discussion to that of Ref.~\cite{arenz2015distinguishing}, any decoherence error that is modeled by quantum channels on the system without explicitly referring to its environment cannot be erased by REAS (as can be seen by considering the case where the depolarizing channel acts on all gates). 

%%%%%%%%%%%%%%%%%%%%%%%%%%%%%%%%%%%%%%%%%%%%%%%%%%%%%%%%%%%%%%%%%%%%

\section{Numerical Results}\label{sec::numericalresults}

We now present numerical simulations that showcase the efficacy of our methods. We consider the setting where $\mcH$ (qubits: $S_0,S_1$) and $\mcH_{\rm env}$ (qubits: $E_0,E_1$) are two-qubit systems (see Fig.~\ref{fig::error_suppression}). The initial (global) state of the circuit is $\ket{0}^{\otimes 4}$ in all simulations. All computational gates and inserted Pauli gates are affected by decoherence noise simulated by applying the following global unitary after each gate $e^{-i\gamma H}\in \mcH\otimes \mcH_{\rm env}$, where $H$ is normalized as $\|H\|_{2}=1$ and the parameter $\gamma$ specifies the scale of noise. The value of $\gamma$ is common across all gates, but the operator $H$ consisting of single-qubit Pauli terms on all systems and two-qubit Pauli terms on $S_0 E_0$ and $S_1 E_1$ that are randomly chosen for each gate (unless specified otherwise). 

In Sec.~\ref{subsec::numericaldeep}, we compare our REAS calibration method to the theoretical prediction for a deep circuit, as well as with conventional calibration techniques (i.e., performing phase estimation without inserting random Pauli gates). In Secs.~\ref{subsec::numericalfirstorder} and \ref{subsec::numericalsinglepauli}, we choose the noise Hamiltonians such that the first-order $\gamma$ term in the shift in rotation angle vanishes to avoid the calibration step, which required too long computation time to give precise results. 

\FloatBarrier

%%%%%%%%%%%%%%%%%%%%%%%%%%%%%%%%%%%%%%%%%%%%%%%%%%%%%%%%%

\subsection{Effectiveness of REAS in Deep Quantum Circuits}\label{subsec::numericaldeep}

\begin{figure}[t]
    \centering
    \includegraphics[width=0.9\linewidth]{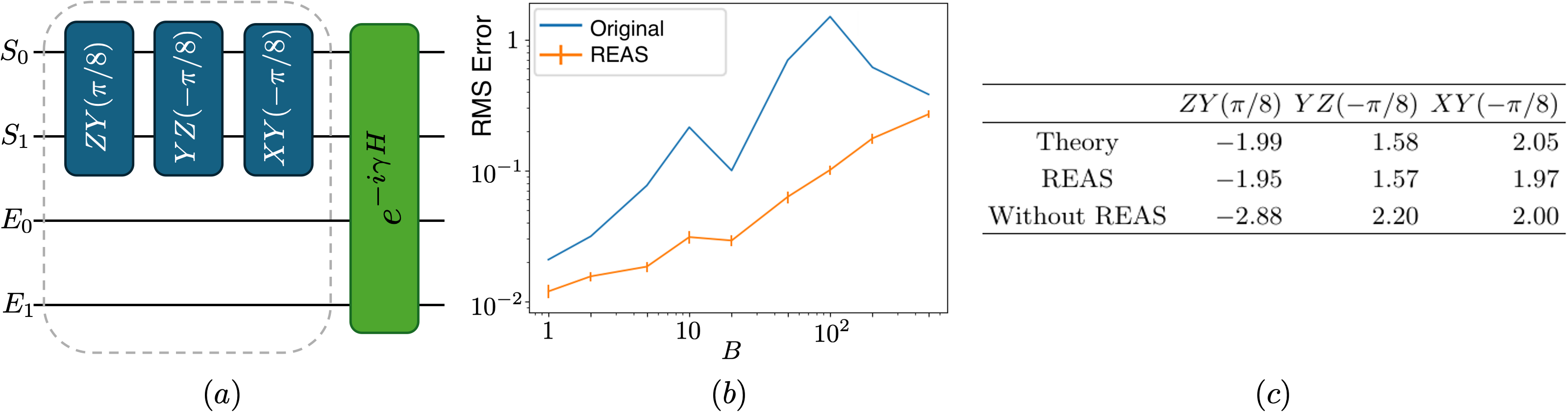}
    \caption{(a) \textit{Deep Circuit.---}We consider a deep circuit composed of repeated blocks of computational gates (blue) $ZY(\pi / 8)$, $YZ(- \pi / 8)$, $XY(- \pi / 8)$, where, e.g., $ZY(\pi / 8)$ corresponds to $e^{-i\frac{\pi}{8}ZY}\otimes I_{\rm env}$ and similar for the others. Following this, decoherence noise is simulated by applying $e^{-i \gamma H}$ (green) on all qubits (including the environment). (b) \textit{REAS Performance in a Deep Circuit.---}The blocks are iterated a number $B$ times, leading to overall circuits of growing depth, and we measure the RMS error with respect to the expectation value of $IZII$ [see Eq.~\eqref{eqn:mse}]. When REAS is applied (orange), the RMS error scales like $B^{0.501}$, which is close to the theoretical upper bound on the scaling of $B^{1/2}$. (c) \textit{Shift in Rotation Angle \textup{(}$10^{-3}$ \textup{rad)}.---}Here we calculate the shift in rotation angle $\Delta(\theta_i)$, i.e., that induced by the coherent noise, in theory as well as with REAS and with conventional calibration techniques, highlighting the performance of the former.
    }
    \label{fig::error_suppression}
\end{figure}

Here, we demonstrate the effectiveness of our REAS scheme in a deep quantum circuit. We consider a circuit block of depth $3$ in Fig.~\ref{fig::error_suppression}(a) containing only two-qubit Pauli computation gates. The string $ZY(\pi / 8)$ indicates the Pauli rotation $e^{-i\frac{\pi}{8}ZY}\otimes I_{\rm env}$; other strings have a similar meaning. In Fig.~\ref{fig::error_suppression}(b), we verify numerically that the theoretical prediction of the RMS between the ideal circuit and a single-shot unitary operator implemented after calibration in a given run of the circuit with random Pauli gates inserted scales as $O(\sqrt{B})$. (Since each block contains three gates, $B=3L$.)

We first perform the REAS calibration to obtain an estimate of the shift in the rotation angle of each gate in $ZY(\pi / 8)$, $YZ(- \pi / 8)$, and $XY(- \pi / 8)$. Throughout the simulation, we fix the noise parameter $\gamma = 0.01$ and introduce a bias in choosing $H$ of the coherent noise $e^{-i\gamma H}$; this is done so that the terms proportional to the appropriate computational Pauli operators are likely to have a large coefficient, in order to clearly demonstrate the effect of calibration. The coherent noise $e^{-i\gamma H}$ invokes modification of rotation angles for $ZY(\theta_1)$, $YZ(\theta_2)$, and $XY(\theta_3)$, i.e., induces the shift $(\theta_1, \theta_2, \theta_3) \rightarrow (\theta_1+\Delta \theta_1, \theta_2 + \Delta \theta_2, \theta_3 + \Delta \theta_3 )$.

We then apply the inverse shift of the rotation angles applied according to the estimates obtained in the REAS calibration, before repeating the circuit block $B \in \{ 1, 2, 5, 10, 20, 50, 100, 200, 500 \}$ times, and measuring the expectation value of the observable $IZII$ (i.e., $Z$ on $S_1$). Finally, we compute the RMS (sampling number: 50) between the ideal expectation value and that obtained in the noisy circuit, both with and without REAS. 

Fig.~\ref{fig::error_suppression}(b) shows that the scaling of the RMS of the difference in the expectation value is nearly of order $B^{1/2}$. Although the theory only predicts that the RMS is upper bounded by a function scaling like $B^{1/2}$---which does not imply that a change in the value of RMS always follows this scaling (e.g., oscillations of the RMS value are possible)---the value of the power calculated here is approximately $0.501$, corresponding closely to the upper bound. In contrast, the corresponding scaling for the case without REAS grows faster (approximately linearly) until it reaches the saturation point. The table in Fig.~\ref{fig::error_suppression}(c) shows that the value of rotation shift $(\theta_1, \theta_2, \theta_3) \rightarrow (\theta_1+\Delta \theta_1, \theta_2 + \Delta \theta_2, \theta_3 + \Delta \theta_3 )$ caused by the coherent noise $e^{-i \gamma H}$ is precisely estimated via REAS calibration with error less than $\times 10^{-4}\ {\rm rad}$. In contrast, the estimate of this rotation shift deviates further from the theoretical prediction when using calibration without inserting random Pauli gates. This result implies that the standard calibration technique based on the phase estimation without inserting random Pauli gates cannot precisely remove the effect of rotation shift and thus the full REAS calibration is required. 

\subsection{Suppression of the First-order Error}\label{subsec::numericalfirstorder}

\begin{figure}
	\centering \hspace{-1em}
	\begin{minipage}{.4\textwidth}
        \centering
        \includegraphics[width=\linewidth]{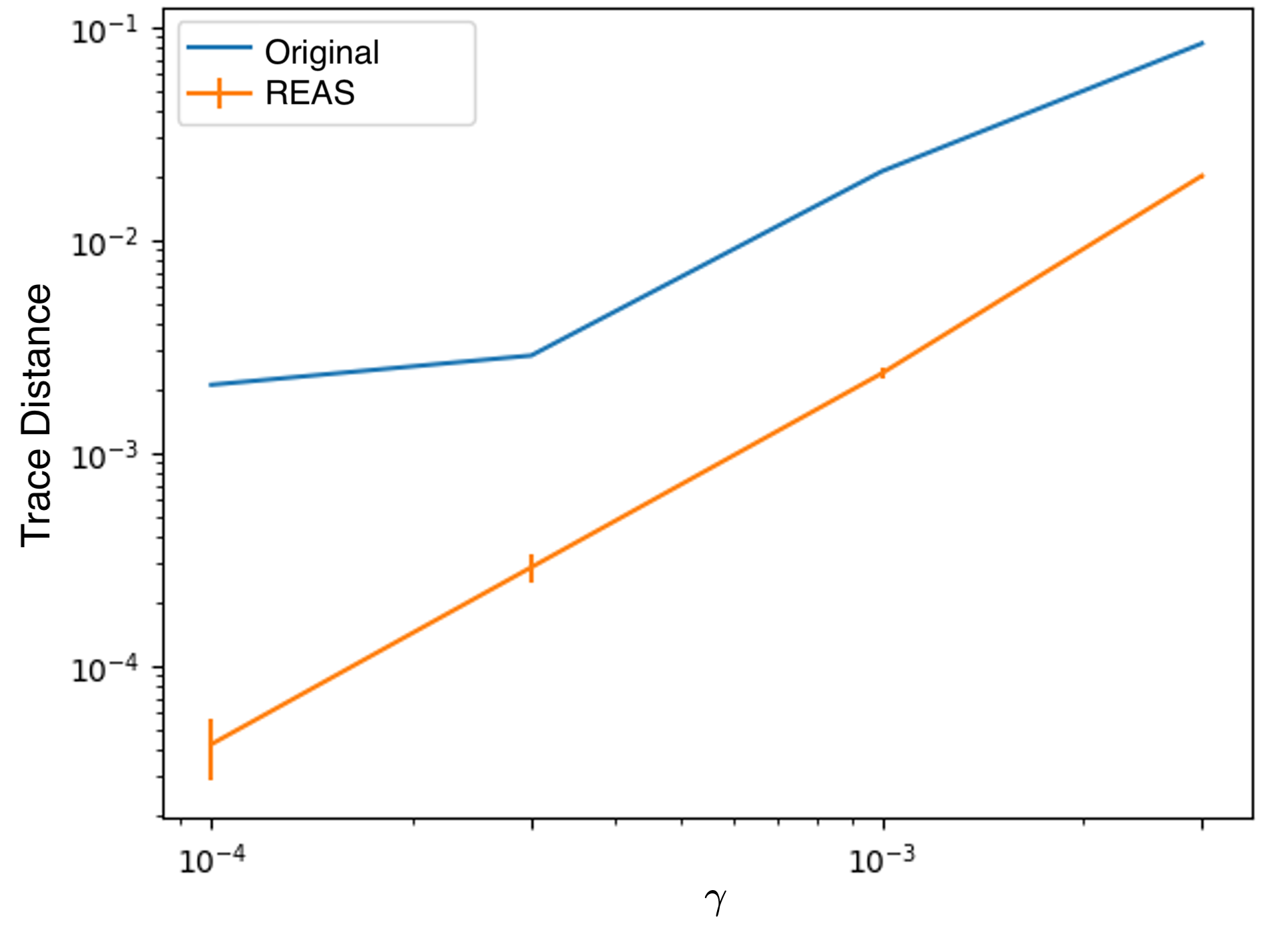}\vspace{-1em}
        \caption{\textit{Suppression of First-order Error.---}Decoherence noise is simulated by applying $e^{-i\gamma H}\in \mcH\otimes \mcH_{\rm env}$ after each gate, where $H$ is normalized as $\|H\|_{2}=1$ and $\gamma$ specifies the scale of noise [see Fig.~\ref{fig::error_suppression}(a)]. In the limit $\gamma\to 0$, the first order error term vanishes when REAS is applied (orange, REAS), in contrast to the case where it is not (blue, Original), as predicted by theory. }
        \label{fig::epsvserr_4000}
	\end{minipage}
	\hspace{0.05\textwidth}
	\begin{minipage}{.4\textwidth}
		\centering\vspace{2em}
        \includegraphics[width=\linewidth]{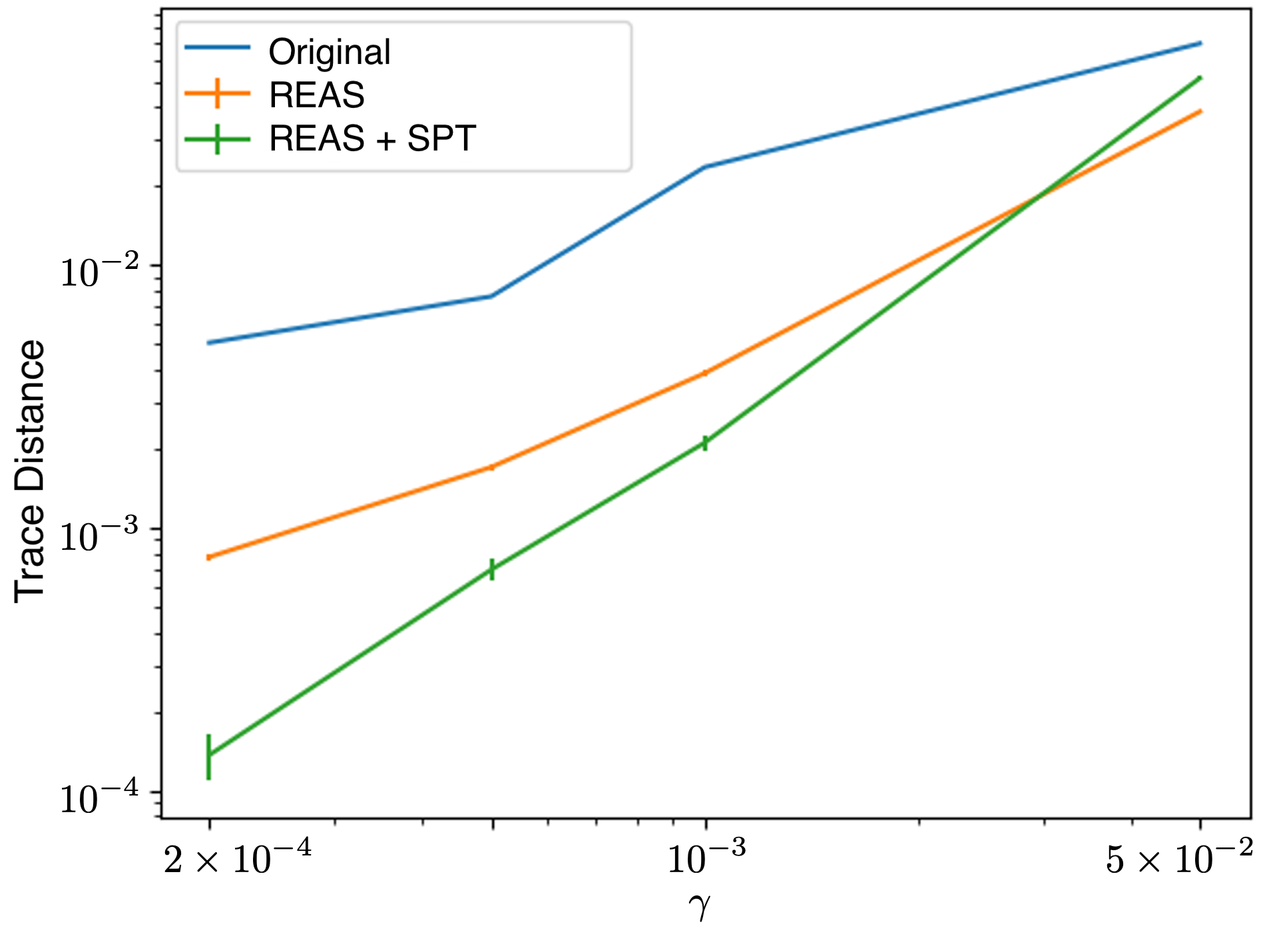}\vspace{-1em}
        \caption{\textit{Effectiveness of Single Pauli Transformation.---}We compare the suppression of decoherence errors in three situations: first, without REAS (blue, Original); second, with REAS but without incorporating the single Pauli transformation procedure (orange, REAS); and third, with REAS and single Pauli transformation (green, REAS + SPT). The behavior improves accordingly in the limit $\gamma\to 0$ (although note that for larger errors, no strict hierarchy applies, as is evident by the crossing of the orange and green lines)}.\vspace{-1em}
        \label{fig::single_transform}
	\end{minipage}
\end{figure}

In Fig.~\ref{fig::epsvserr_4000}, we demonstrate that the first-order term of the error vanishes in the regime where $\gamma$ is near zero (i.e., the weak-coupling regime), as predicted by theory. To do so, we randomly generate a circuit with a depth of 1000 that only contains two-qubit Pauli rotations as the computational gates. For simplicity, we assume that the rotation angles in these random circuits take values in $\pm \tfrac{\pi}{2}$, $\pm \tfrac{\pi}{4}$, and $\pm \tfrac{\pi}{8}$. For the noise parameter $\gamma \in \{0.0001, 0.0003, 0.001, 0.003 \}$, we compute the average trace distance between the state obtained after running the error-free circuit and that obtained after running a noisy circuit both with and without the REAS procedure applied (sampling number: 4000). The simulation in Fig.~\ref{fig::epsvserr_4000} shows a near quadratic dependence of the trace distance on $\gamma$ when using REAS, whereas the scaling for the case without REAS is near linear for small $\gamma$.

\subsection{Effectiveness of the Single Pauli Transformation}\label{subsec::numericalsinglepauli}

In Fig.~\ref{fig::single_transform}, we demonstrate how incorporating the single Pauli transformation helps to reduce decoherence noise. We randomly generate a circuit with a depth of 1000 that contains both single-qubit and two-qubit Pauli gates. For $\gamma \in \{ 0.0002, 0.0005, 0.001, 0.005\}$, we then compute the average trace distance between the state obtained after running the ideal circuit and that obtained after running the noisy circuit, both with and without REAS applied and both with and without the single Pauli transformation (sampling number: 4000). The simulation shown in Fig.~\ref{fig::single_transform} shows that the circuit implemented with the single Pauli transformation included displays a smaller trace distance in the small $\gamma$ regime compared to the case where Pauli rotation is implemented directly without said transformation procedure.

%%%%%%%%%%%%%%%%%%%%%%%%%%%%%%%%%%%%%%%%%%%%%%%%%%%%%%%%%%%%%%%%%%

\section{Experimental Demonstration}\label{sec::experiments}

To complement the numerical results above, we finally demonstrate the performance of a special case of our REAS method on available quantum computing systems. For this experiment, we make the following assumptions which seem well-justified in modern quantum computers with superconducting qubits: i) two-qubit gates have much larger errors than single-qubit ones, the latter of which are negligible; and ii) standard calibration (i.e., procedures used in IBM Quantum systems) of rotation angles for two-qubit Pauli rotation gates is highly accurate. Based on these assumptions, we tested a variant of REAS using the following setup: i) the single Pauli transformation step was not applied, meaning only coherent errors were addressed; ii) ordinary calibration (without twirling) was performed for all gates; iii) consecutive single-qubit gates applied to the same qubit were merged as much as possible (i.e., in Fig.~\ref{fig::reas_schematic}, the inserted gates were merged with the computation gates). As such, the experimental demonstration does not unleash the full power of REAS and instead performs a variant of randomized compiling~\cite{wallman2016noise, layden2023quantum, santos2024deeper}. Nonetheless, our experiment shows that such a method works well for suppressing the accumulation of errors in quantum circuits with many qubits, which---to the best of our knowledge---has not yet been demonstrated on presently available platforms.

We use Trotterized circuits to simulate the time evolution of a one-dimensional transverse field Ising model and analyze how the error scales with respect to the number of Trotter steps, which is proportional to the number of 2-qubit gate layers. The Hamiltonian of the model is 
\begin{equation}\label{eqn::isinghamilatonian}
    H = -\sum_{i=1}^{N-1}{g_{i} Z_i Z_{i+1}} -\sum_{i=1}^{N}{h_{i} X_i}
\end{equation}
where $g_i$ represents the interaction strength between spins $i$ and $i+1$, $h_i$ represents the external magnetic field strength at site $i$, and $N$ is the number of spins (i.e. qubits) of the system.

In order to calculate the ideal output of the circuit, we use the so-called compute-uncompute procedure, which consists of a first stage where the original circuit $U$ is applied, and a second stage where it is reversed via application of $U^\dagger$. The ideal output state of such circuits applied to an initial blank register $|0\rangle^{\otimes N}$ will be identical to the input state, which has trivial expectation value (i.e., either 0 or 1) for any Pauli observable. This allows us to compute how far the experimentally obtained expectation values are from the ideal ones.

\begin{figure}[t]
    \centering
    \includegraphics[width=0.8\linewidth]{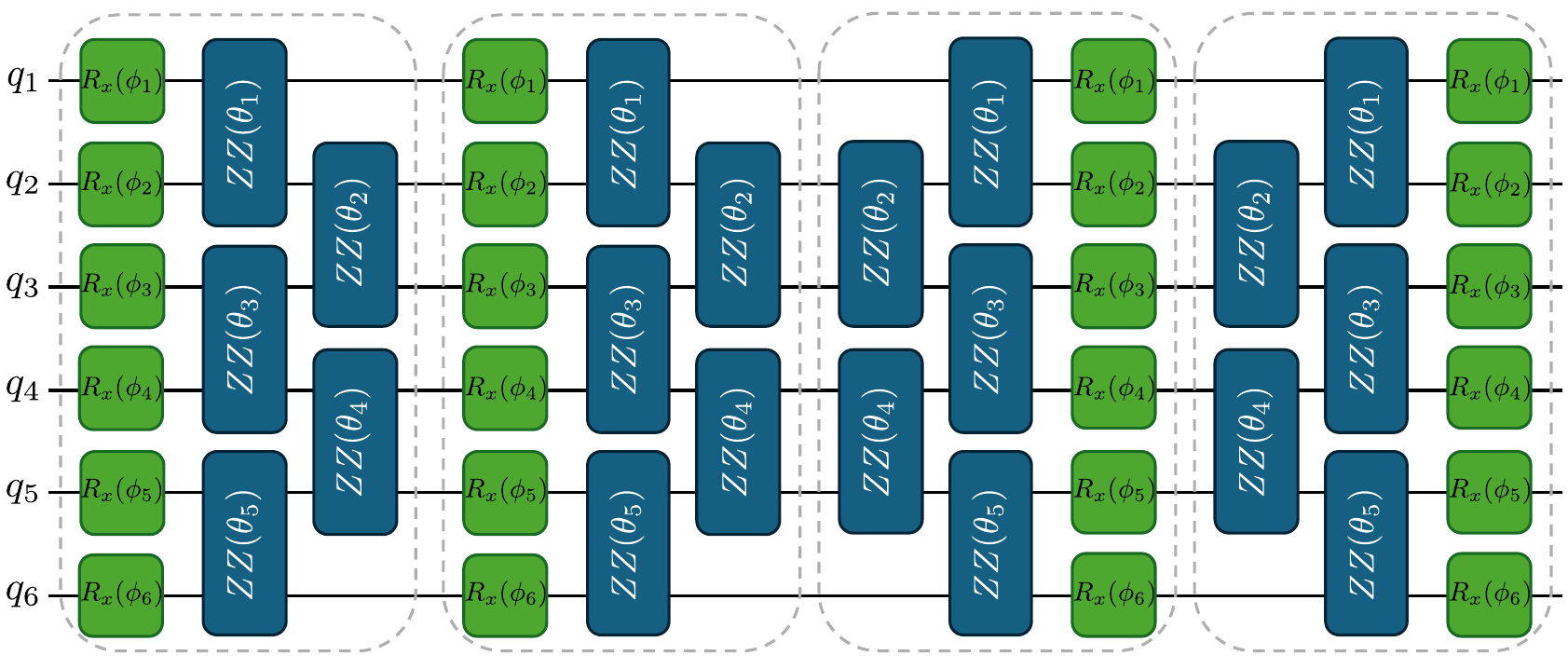}
    \caption{\textit{Example Circuit.---}A compute-uncompute Trotterized circuit of simulating the time evolution of a one-dimensional transversal field Ising model on six qubits using two Trotter steps. For the experimental demonstration, we employ circuits with the same structure on 100 qubits with 1 to 16 Trotter steps.}
    \label{fig::exp_circuit}
\end{figure}

In Fig.~\ref{fig::exp_circuit}, we show an example circuit in the case of $N=6$ with two Trotter steps. As shown in the figure, the $i^{\text{th}}$ $ZZ$ term in the Hamiltonian Eq.~\eqref{eqn::isinghamilatonian} generates a $ZZ(\theta_i)$ rotation gate in each Trotter step, where $\theta_i$ is an angle proportional to $g_i$ and the time step. In our experiment, we choose the value $\theta_i$ by sampling the pulse amplitude that implements the $ZZ(\theta_i)$ rotation gate (see App.~\ref{app::experiments} for details). As for the $X_i$ term in the Hamiltonian, this evolution is implemented by an $R_x(\phi_i)$ gate, where $\phi_i$ is an angle proportional to $h_i$ and the time step. We use the uniform value $\phi_i=0.1 \pi$ for all $i=1\ldots N$ in our experiment.

We evaluate the amount of errors by the RMS error of the expectation value of a Pauli observable $P$, i.e.,
\begin{equation}\label{eqn:mse}
\sqrt{
\sum_k{p_k \|\braket{P}_{\mathrm{ideal}} - \braket{P}_{\mathrm{exp}(k)}\|^2}
}
\end{equation}
where $p_k$ is the probability of drawing the $k^{\text{th}}$ set of random Paulis in the REAS procedure and $\braket{P}_{\mathrm{exp}(k)}$ represents the expectation value of $P$ experimentally observed in the $k^{\text{th}}$ circuit realization.\footnote{Note that we are interested in the average of errors of expectation values (i.e., the effect of error suppression), not in measuring the error on the averaged expectation value (i.e., the effect of error mitigation).}

\begin{figure}[t]
    \centering
    \includegraphics[width=\linewidth]{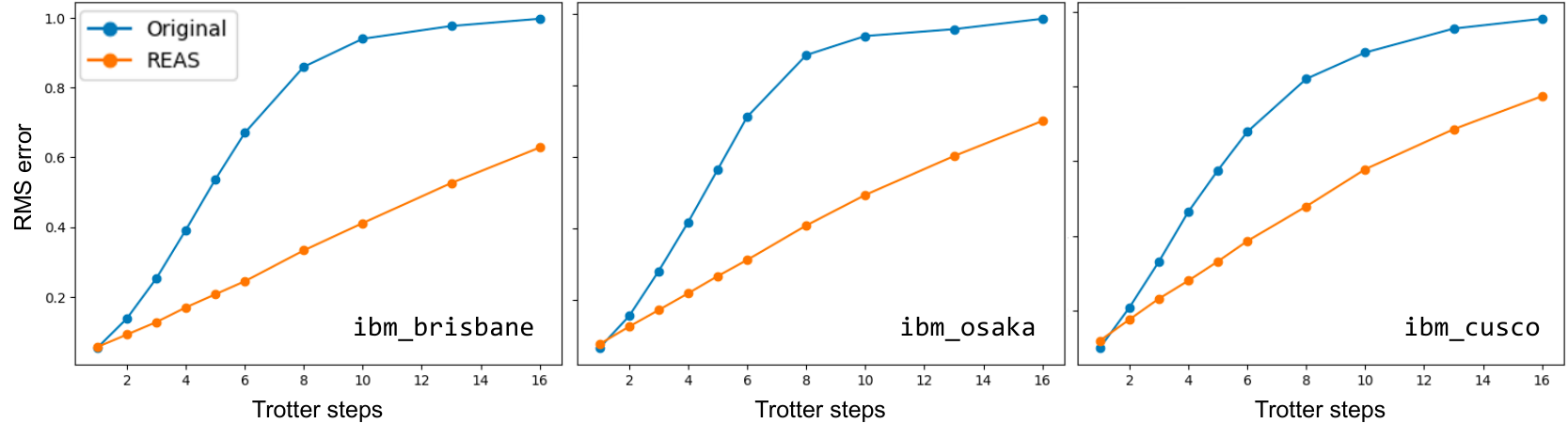}
    \caption{\textit{Experimental Demonstration.---}Average RMS error over 100 weight-1 Z observables by Trotter steps with REAS (orange, REAS) and without REAS (blue, Original) using three IBM Quantum processors: \texttt{ibm\_brisbane}, \texttt{ibm\_osaka} and \texttt{ibm\_cusco}. These results highlight the efficacy and robustness of our method across distinct quantum devices.
    }
    \label{fig::main_exp_result}
\end{figure}

For the experiment, we made use of 127-qubit IBM Quantum processors~\cite{IBMQuantum}, namely \texttt{ibm\_brisbane}, \texttt{ibm\_osaka} and \texttt{ibm\_cusco}, comparing the average error in expectation values both with and without the REAS procedure applied. In both cases, we considered Trotter steps: 1, 2, 3, 4, 5, 6, 8, 10, 13, 16. Here, e.g., 16 Trotter steps corresponds to 64 two-qubit gate layers, since each step consists of two layers and the steps are doubled by uncomputing. For the cases where REAS is applied, we sampled 25 circuits for each number of Trotter steps, meaning that we ran ten circuits for the situation without REAS and 250 circuits for that with it. In order to balance the total number of shots, we set 10000 shots for the former case and 400 shots for the latter. For each circuit, we measured all weight-1 Pauli observables for 100 qubits and computed the RMS error [see Eq.~\eqref{eqn:mse}] for each observable. To accurately estimate the expectation values of the observables, we applied a readout error mitigation technique called Twirled Readout Error eXtinction (TREX)~\cite{van2022model}. Figure~\ref{fig::main_exp_result} demonstrates how the RMS error (averaged over all 100 weight-1 $Z$ observables) scales with Trotter steps. Clearly, the suppression of errors via REAS improves as the number of Trotter steps grows for all experiments on the three processors. 

%%%%%%%%%%%%%%%%%%%%%%%%%%%%%%%%%%%%%%%%%%%%%%%%%%%%%%%%%%%%%%%%%%

\section{Discussion}\label{sec::summary}

We have presented an advanced, robust method of quantum error suppression. The ability to deal with non-Clifford circuits and incorporate both coherent and decoherence errors, both in the original circuit and in its modification, stems from combining robust phase estimation techniques with a calibrated Pauli twirling-type procedure, as described in Secs.~\ref{sec::reas} and~\ref{sec::results}. To highlight the immediate applicability of our results, we present numerical evidence for the quadratic reduction in error growth presented in Sec.~\ref{sec::numericalresults} and furthermore present an experimental demonstration in Sec.~\ref{sec::experiments}. Since the types of errors that our method is tailor-made to deal with go far beyond the current state-of-the-art, we expect our methods to be widely adopted across a variety of platforms.

Looking forward, extending our method to deal with leakage/qubit loss will be an important generalization. Secondly, an improved technique of dealing with correlated errors across many qubits, e.g., non-local crosstalk. Lastly, actively suppressing errors in complex non-Markovian processes remains of utmost importance, especially as the timescales over which one can access quantum systems are ever shortening, making non-trivial temporal correlations more prevalent~\cite{white2020demonstration,guo2021experimental,goswami2021experimental,figueroa2021randomized,figueroa2022towards,white2022nonmarkovian,white2023filtering}. \vspace{-1em}

\section*{Note}
Upon completion of this work, we became aware of a related method that similarly generalizes Pauli twirling techniques to suppress errors in non-Clifford circuits~\cite{santos2024deeper}.

\begin{acknowledgments}
We thank Joseph Emerson for interesting discussions. We acknowledge the use of IBM Quantum services for this work. The views expressed are those of the authors, and do not reflect the official policy or position of IBM or the IBM Quantum team. P.T. acknowledges funding from the Japan Society for the Promotion of Science (JSPS) Postdoctoral Fellowships for Research in Japan and the IBM-UTokyo Laboratory. N.Y. is supported by PRESTO, JST, Grant No.\,JPMJPR2119, COI-NEXT program Grant No. JPMJPF2221, JST ERATO Grant Number JPMJER2302, and JST CREST Grant Number JPMJCR23I4, Japan and IBM Quantum. M.M. was supported by MEXT Quantum Leap Flagship Program (MEXT QLEAP) JPMXS0118069605, JPMXS0120351339, JSPS KAKENHI Grant No. 21H03394 and No. 23K21643, and IBM Quantum. 
\end{acknowledgments}

%\bibliography{references}


\newpage

\appendix

\vspace*{8mm}
\pdfbookmark[0]{Appendices}{Appendices}
%\vspace*{5mm}
\label{supplementalmaterial}
\phantomsection
\begin{center}
\begin{LARGE}
APPENDICES
\end{LARGE}
\end{center}

\section{Suppression of Coherent Errors by REAS}\label{app::reascalibration}

Here, we demonstrate how the REAS protocol can deal with coherent errors. For simplicity, suppose that the gate sequence that one wishes to implement in the computation does not involve parallel application of Pauli rotation gates, i.e., it can be written as
\begin{align}\label{eq::no_parallel}
    \prod_{j=1}^L e^{-i\sigma_j\theta_j}\in \mcL(\mcH)
\end{align}
using single- or two-qubit Pauli operators $\sigma_1,\ldots ,\sigma_L$ and rotation angles $\theta_1,\ldots ,\theta_N$. Again, we abbreviate the tensor product with the identity on the systems that undergo trivial dynamics. Suppose that coherent errors affect each Pauli rotation $e^{-is\sigma_j\theta_j}\ (s\in \{+1,-1\})$ so that the actual noisy gate implemented is expressed as $e^{-is\sigma_j\theta_j}[I+i\epsilon C_{s,j}]$ for some unitary $I+i\epsilon C_{s,j}$ where $\epsilon>0$ and $\|C_{s,j}\|_{\rm op}\leq 1$ for all $(s,j)$. 
Here, we assume that the same error occurs on the same type of gate at any time, namely, $C_{s,j}=C_{s,k}$ if $(\sigma_j,\theta_j)=(\sigma_k,\theta_k)$, which is rather natural and required to enable calibration. 

Since each $C_{s,j}$ is defined on $\mcH$, it can be decomposed in terms of a linear combination of $n$-qubit Pauli operators as
\begin{align}
    C_{s,j}=:\sum_{\vec{v}\in \{0,1,2,3\}^n}
    \alpha_{\vec{v},j}(s)\sigma_{\vec{v}}
\end{align}
using complex numbers $\alpha_{\vec{v}}(s) \in \mathbb{C}$. Importantly, note that $\alpha_{\vec{v},j}(s)$ can be further decomposed into a sum of $s$-dependent part and $s$-independent part as
$\alpha_{\vec{v},j}(s)=:\overline{\alpha}_{\vec{v},j}+s\Delta\alpha_{\vec{v},j}$, where $\overline{\alpha}_{\vec{v},j}:=\tfrac{1}{2}[\alpha_{\vec{v},j}(1)+\alpha_{\vec{v},j}(-1)]$ and $\Delta {\alpha}_{\vec{v},j}:=\tfrac{1}{2}[\alpha_{\vec{v},j}(1)-\alpha_{\vec{v},j}(-1)]$. We will frequently make use of this technique throughout this analysis. 

Since the operator $[I+i\epsilon C_{s,j}]$ is unitary, it follows that
\begin{align}
    \left[
    I+i\epsilon
    \sum_{\vec{v}}
    \alpha_{\vec{v},j}(s)
    \sigma_{\vec{v}}
    \right]
    \left[
    I-i\epsilon
    \sum_{\vec{v}}
    \alpha^*_{\vec{v},j}(s)
    \sigma_{\vec{v}}
    \right] &= I+i\epsilon\left(
    \sum_{\vec{v}}
    (\alpha_{\vec{v},j}(s)-\alpha^*_{\vec{v},j}(s))
    \right)
    +O(\epsilon^2)
    =I
    ,
\end{align}
which implies that the 0$^\text{th}$ order $\epsilon$ term $\alpha_{\vec{v},j}^{(0)}$ of $\alpha_{\vec{v},j}$ is real. 
Moreover, we can take the 0$^\text{th}$ order $\epsilon$ term $\alpha_{\vec{0},j}$, namely $\alpha_{\vec{0},j}^{(0)}$, the dominant term of the coefficient of $I$, to be $0$ without loss of generality. This is because it can be canceled via an appropriate global phase, as follows
\begin{align}
    \frac{1-i\epsilon \alpha_{\vec{0},j}^{(0)}}{\sqrt{1+(\epsilon \alpha_{\vec{0},j}^{(0)})^2}}
    \left[
    I+i\epsilon
    \sum_{\vec{v}}
    (\alpha_{\vec{v},j}^{(0)}(s)+O(\epsilon))
    \sigma_{\vec{v}}
    \right]
    &=
    \frac{1}{\sqrt{1+(\epsilon \alpha_{\vec{0},j}^{(0)})^2}}
    \left[
    I+i\epsilon\left\{
    O(\epsilon)I+
    \sum_{\vec{v}\neq \vec{0}}
    \left(\alpha_{\vec{v},j}^{(0)}(s)
    +O(\epsilon)
    \right)\sigma_{\vec{v}}
    \right\}
    \right]
    \nonumber\\
    &=
    (1-O(\epsilon^2))
    \left[
    I+i\epsilon\left\{
    O(\epsilon)I+
    \sum_{\vec{v}\neq \vec{0}}
    \left(\alpha_{\vec{v},j}^{(0)}(s)
    +O(\epsilon)
    \right)\sigma_{\vec{v}}
    \right\}
    \right]
    \nonumber\\
    &=:
    I+i\epsilon
    \sum_{\vec{v}\neq \vec{0}}{\alpha}_{\vec{v},j}^{(0)}(s)\sigma_{\vec{v}}
    +O(\epsilon^2) .
\end{align}
Thus, we assume that $C_{s,j}$ is defined such that $\alpha_{\vec{0},j}^{(0)}(s)=0$ for all $(s,j)$.

Now, we move on to the error analysis of the gate sequence obtained by inserting random Pauli gates via REAS into Eq.~(\ref{eq::no_parallel}). This leads to
\begin{align}
    \sigma_{\vec{v}_L}e^{-is_L\sigma_L\theta_L}
    \prod_{j=1}^{L-1}\left[
    \sigma_{\vec{v}_{j+1:j}}
    e^{-is_j\sigma_j\theta_j}\right]
    \sigma_{\vec{v}_1},
\end{align}
where $\vec{v}_1,\ldots ,\vec{v}_L$ are uniformly randomly chosen vectors. For simplicity, here we have assumed that a random Pauli gate is inserted between all single Pauli rotation gates. Recall that $s_j\in \{+1,-1\}$ is defined as
\begin{align}
    s_j:=
    \begin{cases}
        1&(\sigma_j\text{ and }\sigma_{\vec{v}_j}\text{ commutes})\\
        -1&(\sigma_j\text{ and }\sigma_{\vec{v}_j}\text{ anti-commutes}).\\
    \end{cases}
\end{align}
When such a gate sequence is applied on an arbitrary input state $\rho\in \mcL(\mcH)$, the result is 
\begin{align}\label{eq::no_para_ideal}
    &\left(
    \frac{1}{4^n}
    \right)^L\sum_{\vec{v}_1,\ldots,\vec{v}_L}
    \left[\sigma_{\vec{v}_L}e^{-is_L\sigma_L\theta_L}
    \overleftarrow{\prod_{j=1}^{L-1}}\left[
    \sigma_{\vec{v}_{j+1:j}}
    e^{-is_j\sigma_j\theta_j}\right]
    \sigma_{\vec{v}_1}\right]\rho
    \left[\sigma_{\vec{v}_1}
    \overrightarrow{\prod_{j=1}^{L-1}}\left[
    e^{is_j\sigma_j\theta_j}
    \sigma_{\vec{v}_{j+1:j}}
    \right]
    e^{is_L\sigma_L\theta_L}
    \sigma_{\vec{v}_L}
    \right]
    \nonumber\\
    =&
    \left(
    \frac{1}{4^n}
    \right)^L\sum_{\vec{v}_1,\ldots,\vec{v}_L}
    \left[\sigma_{\vec{v}_L}e^{-is_L\sigma_L\theta_L}
    \overleftarrow{\prod_{j=1}^{L-1}}\left[
    \sigma_{\vec{v}_{j+1}}\sigma_{\vec{v}_j}
    e^{-is_j\sigma_j\theta_j}\right]
    \sigma_{\vec{v}_1}\right]\rho
    \left[\sigma_{\vec{v}_1}
    \overrightarrow{\prod_{j=1}^{L-1}}\left[
    e^{is_j\sigma_j\theta_j}\sigma_{\vec{v}_{j}}\sigma_{\vec{v}_{j+1}}
    \right]
    e^{is_L\sigma_L\theta_L}
    \sigma_{\vec{v}_L}
    \right]
    .
\end{align}
Here, we use notation $\overleftarrow{\prod}$ and $\overrightarrow{\prod}$ to distinguish the order of applying the gate sequences, defined such that 
\begin{align}
    \overleftarrow{\prod_{j=1}^L}M_j:=M_L\cdots M_1 \quad\quad\quad\quad\text{and}\quad\quad\quad\quad
    \overrightarrow{\prod_{j=1}^L}M_j:=M_1\cdots M_L
\end{align}
for linear operators $M_1,\ldots,M_L$. It is straightforward to verify that when no error terms appears, Eq.~(\ref{eq::no_para_ideal}) reduces to the ideal state transformation
\begin{align}\label{eq::no_para_0th}
    \left(\overleftarrow{\prod_{j=1}^L}e^{-i\sigma_j\theta_j}\right)\rho
    \left(\overrightarrow{\prod_{j=1}^L}e^{i\sigma_j\theta_j}\right),
\end{align}
which can be seen by moving the inner Pauli operators $\sigma_{\vec{v}_j}$ outward, making them pass through $e^{\pm is_j\sigma_j\theta_j}$ and thereby changing $e^{\pm is_j\sigma_j\theta_j}$ to $e^{\pm i\sigma_j\theta_j}$, and subsequently canceling with the corresponding outer $\sigma_{\vec{v}_j}$ operator.

We now prove that, as claimed in the main text, coherent errors lead to Eq.~(\ref{eq::no_para_ideal}) being transformed as
\begin{align}\label{eq::no_para_wanttoshow}
    \left(\overleftarrow{\prod_{j=1}^L}e^{-i\sigma_j(\theta_j-\epsilon \Delta\theta_j)}\right)\rho
    \left(\overrightarrow{\prod_{j=1}^L}e^{i\sigma_j(\theta_j-\epsilon \Delta\theta_j)}\right)
    +O(\epsilon+\epsilon^2 L),
\end{align}
where $\epsilon\Delta\theta_j:=\tan^{-1}(\epsilon \Delta\alpha_{\vec{w}_{j},j}^{(0)})$ where $\sigma_{\vec{w}_j}:=\sigma_{j}$. Our technique of suppressing the accumulation of errors is based on the following equations regarding averaging over an $n$-qubit Pauli operator $\sigma$
\begin{align}\label{eq::randompauliaverageidentity}
    \left(\frac{1}{4^n}
    \right)
    \sum_{\vec{v}\in\{0,1,2,3\}^n}
    \sigma_{\vec{v}}
    \sigma
    \sigma_{\vec{v}}
    =\begin{cases}
        I& \quad\quad \text{for }\sigma=I\\
        0& \quad\quad \text{for }\sigma\neq I
    \end{cases}
\end{align}
and 
\begin{align}\label{eq::no_para_s}
    \left(\frac{1}{4^n}
    \right)
    \sum_{\vec{v}\in\{0,1,2,3\}^n}
    \sigma_{\vec{v}}
    s(\sigma_{\vec{v}},\sigma')\sigma
    \sigma_{\vec{v}}=
    \begin{cases}
        \sigma'& \quad\quad \text{for }\sigma=\sigma'\\
        0& \quad\quad \text{for }\sigma\neq\sigma'
    \end{cases}
\end{align}
where $s(\sigma_{\vec{v}},\sigma')$ is defined as
\begin{align}
    s(\sigma_{\vec{v}},\sigma'):=
    \begin{cases}
        1&(\sigma_{\vec{v}}\text{ and }\sigma'\text{ commutes})\\
        -1&(\sigma_{\vec{v}}\text{ and }\sigma'\text{ anti-commutes}).
    \end{cases}
\end{align}
With this, the error term $i\epsilon C_{s_{j'}, j'}$ transforms as
\begin{align}
    \frac{1}{4^n}\sum_{\vec{v}_{j'}}\sigma_{\vec{v}_{j'}}
    i\epsilon C_{s_{j'}, j'}
    \sigma_{\vec{v}_{j'}}=
    \frac{1}{4^n}\sum_{\vec{v}_{j'}}\sigma_{\vec{v}_{j'}}
    \left[\sum_{\vec{u}}i\epsilon(\overline{\alpha}_{\vec{u},j'}+s_{j'}\Delta\alpha_{\vec{u},j'})
    \sigma_{\vec{u}}\right]\sigma_{\vec{v}_{j'}}
    =
    O(\epsilon^2)I+i\epsilon\Delta\alpha_{\vec{w}_{j'},j'}\sigma_{j'}
\end{align}
where $\sigma_{\vec{w}_{j'}}=\sigma_{j'}$ by definition. This expression can be simplified by absorbing the first-order $\epsilon$ part in $i\epsilon s_{j'}\Delta\alpha_{\vec{w}_{j'},j'}\sigma_{j'}$
to a shift in the rotation angle of $e^{-is_{j'}\sigma_{j'}\theta_{j'}}$ as follows
\begin{align}
    e^{-is_{j'}\sigma_{j'}\theta_{j'}}[I+i\epsilon C_{s_{j'},j'}]
    &=:g\left(
    e^{-is_{j'}\sigma_{j'}\theta_{j'}}
    \right)\nonumber\\
    &=e^{-is_{j'}\sigma_{j'}\theta_{j'}}
    e^{i\epsilon s_{j'} \Delta\theta_{j'}\sigma_{j'}}
    \frac{1-is_{j'}\epsilon \Delta\alpha_{\vec{w}_{j'},j'}^{(0)}}{\sqrt{1+(\epsilon\Delta\alpha_{\vec{w}_{j'},j'}^{(0)})^2}}
    \left[
    I+i\epsilon 
    \sum_{\vec{u}\neq \vec{0}}
    (\overline{\alpha}_{\vec{u},j'}+s_{j'}\Delta\alpha_{\vec{u},j'})\sigma_{\vec{u}}
    +O(\epsilon^2)
    \right]
    \nonumber\\
    &=
    e^{-is_{j'}\sigma_{j'}(\theta_{j'}-\epsilon \Delta\theta_{j'})}
    \left[1-O(\epsilon^2)\right]
    \left[
    I+i\epsilon\sum_{\vec{u}\notin \{\vec{0},\vec{w}_{j'}\}}
    (\overline{\alpha}_{\vec{u},j'}+s_{j'}\Delta\alpha_{\vec{u},j'})\sigma_{\vec{u}}
    +i\epsilon \overline{\alpha}_{\vec{w}_{j'},j'}\sigma_{j'}
    +O(\epsilon^2)
    \right]
    \nonumber\\
    &=
    e^{-is_{j'}\sigma_{j'}(\theta_{j'}-\epsilon \Delta\theta_{j'})}
    \left[
    I+i\epsilon\sum_{\vec{u}\notin \{\vec{0},\vec{w}_{j'}\}}
    (\overline{\alpha}_{\vec{u},j'}+s_{j'}\Delta\alpha_{\vec{u},j'})\sigma_{\vec{u}}
    +i\epsilon \overline{\alpha}_{\vec{w}_{j'},j'}\sigma_{j'}
    +O(\epsilon^2)
    \right]
    \nonumber\\
    &=:e^{-is_{j'}\sigma_{j'}(\theta_{j'}-\epsilon \Delta\theta_{j'})}[I+i\epsilon D_{s_{j'},j'}]
    =:g'\left(
    e^{-is_{j'}\sigma_{j'}(\theta_{j'}-\epsilon \Delta\theta_{j'})}
    \right)
    .
\end{align}
Thus, the error term $i\epsilon D_{s_{j'},j'}$ no longer contains the first-order $\epsilon$ term proportional to $s_{j'}\sigma_{{j'}}$. Using this result, Eq.~(\ref{eq::no_para_ideal}) with coherent errors can be rewritten as
\begin{align}\label{eq::no_para_gg'}
    &\left(
    \frac{1}{4^n}
    \right)^L\sum_{\vec{v}_1,\ldots,\vec{v}_L}
    \left[g(\sigma_{\vec{v}_L})g(e^{-is_L\sigma_L\theta_L})
    \overleftarrow{\prod_{j=1}^{L-1}}\left[
    g(\sigma_{\vec{v}_{j+1:j}})
    g(e^{-is_j\sigma_j\theta_j})\right]
    g(\sigma_{\vec{v}_1})\right]\rho
    \left[\cdots
    \right]
    \nonumber\\
    =&\left(
    \frac{1}{4^n}
    \right)^L\sum_{\vec{v}_1,\ldots,\vec{v}_L}
    \left[g(\sigma_{\vec{v}_L})g'(e^{-is_L\sigma_L(\theta_L-\epsilon \Delta\theta_L)})
    \overleftarrow{\prod_{j=1}^{L-1}}\left[
    g(\sigma_{\vec{v}_{j+1:j}})
    g'(e^{-is_j\sigma_j(\theta_j-\Delta\theta_j)})\right]
    g(\sigma_{\vec{v}_1})\right]
    \rho
    \left[\cdots
    \right]
\end{align}
where the error function $g$ on inserted random Pauli operators $\sigma_{\vec{v}}$ is defined as
$g(\sigma_{\vec{v}})=:\sigma_{\vec{v}}[i+i\epsilon \widetilde{C}_{\vec{v}}]$ such that $i\epsilon \widetilde{C}_{\vec{v}}$ has no $O(\epsilon)$ term proportional to $I$. Here, the notation [$\hdots$] on the right of $\rho$ represents the Hermitian conjugate of the corresponding product on the left; we will continue to use similar notation below.

We move to analyze the suppression of errors. Consider a single error term on a particular Pauli rotation gate $e^{-is_{j'}\sigma_{j'}\theta_{j'}}$ in the computation, which leads to an expression of the form~\footnote{We will only explicitly show the calculation for the error term $i\epsilon D_{s_{j'}, j'}$ in question appearing on the left of $\rho$; since the term on the right is simply its Hermitian conjugate, the analysis there is similar.}
\begin{align}\label{eq::no_para_one_error}
    &\left(
    \frac{1}{4^n}
    \right)^L\sum_{\vec{v}_1,\ldots,\vec{v}_L}
    \left[
    \sigma_{\vec{v}_L}
    [\cdots ]
    \sigma_{\vec{v}_{j'+1,j'}}
    (e^{-is_{j'}\sigma_{j'}(\theta_{j'}-\epsilon \Delta\theta_{j'})}i\epsilon D_{s_{j'},j'})
    \sigma_{\vec{v}_{j'-1:j'}}
    [\cdots]
    \sigma_{\vec{v}_1}\right]
    \rho 
    \left[\sigma_{\vec{v}_1}
    [\cdots]
    \sigma_{\vec{v}_L}\right] 
    \nonumber\\
    =&
    \left(
    \frac{1}{4^n}
    \right)^L\sum_{\vec{v}_1,\ldots,\vec{v}_L}
    \left[
    \sigma_{\vec{v}_L}
    [\cdots ]
    \sigma_{\vec{v}_{j'+1}}\sigma_{\vec{v}_{j'}}
    (e^{-is_{j'}\sigma_{j'}(\theta_{j'}-\epsilon \Delta\theta_{j'})}i\epsilon D_{s_{j'},j'})
    \sigma_{\vec{v}_{j'}}\sigma_{\vec{v}_{j'-1}}
    [\cdots]
    \sigma_{\vec{v}_1}\right]
    \rho 
    \left[\sigma_{\vec{v}_1}
    [\cdots]
    \sigma_{\vec{v}_L}\right] 
    .
\end{align}
All pairs of inserted Pauli operators $\sigma_{\vec{v}_j}$ other than $\sigma_{\vec{v}_{j'}}$ that sandwich $i\epsilon D_{s_{j'},j'}$ can pass through the operator $e^{\pm is_j\sigma_j(\theta_j-\epsilon \Delta\theta_j)}$ and cancel with each other. Thus, Eq.~(\ref{eq::no_para_one_error}) can be rewritten as
\begin{align}
    &\frac{1}{4}\sum_{\vec{v}_{j'}}
    \left[\left(
    \overleftarrow{\prod_{j=j'+1}^L}e^{-i\sigma_{j}(\theta_j-\epsilon\Delta\theta_j)}\right)\sigma_{\vec{v}_{j'}}e^{-is_{j'}\sigma_{j'}(\theta_{j'}-\epsilon\Delta\theta_{j'})}i\epsilon
    D_{s_{j'},j'}\sigma_{\vec{v}_{j'}}\left(
    \overleftarrow{\prod_{j=1}^{j'-1}}e^{-i\sigma_{j}(\theta_j-\epsilon\Delta\theta_j)}\right)\right]
    \rho
    \left[
    \overrightarrow{\prod_{j=1}^L}e^{is_j\sigma_j(\theta_j-\epsilon\Delta\theta_j)}
    \right]
    \nonumber\\
    =&\left[\left(
    \overleftarrow{\prod_{j=j'+1}^L}e^{-i\sigma_{j}\theta_j}\right)e^{-i\sigma_{j'}(\theta_{j'}-\epsilon\Delta\theta_{j'})}\left(
    \frac{1}{4^n}\sum_{\vec{v}_{j'}}
    \sigma_{\vec{v}_{j'}}i\epsilon
    D_{s_{j'},j'}\sigma_{\vec{v}_{j'}}\right)\left(
    \overleftarrow{\prod_{j=1}^{j'-1}}e^{-i\sigma_{j}(\theta_j-\epsilon\Delta\theta_j)}\right)\right]
    \rho
    \left[
    \overrightarrow{\prod_{j=1}^L}e^{is_j\sigma_j(\theta_j-\epsilon\Delta\theta_j)}
    \right]
    \nonumber\\
    =&
    \left[\left(
    \overleftarrow{\prod_{j=j'+1}^L}e^{-i\sigma_{j}\theta_j}\right)e^{-i\sigma_{j'}(\theta_{j'}-\epsilon\Delta\theta_{j'})}\left(
    O(\epsilon^2)I+O(\epsilon^2)\sigma_{\vec{w}_{j'}}\right)\left(
    \overleftarrow{\prod_{j=1}^{j'-1}}e^{-i\sigma_{j}(\theta_j-\epsilon\Delta\theta_j)}\right)\right]
    \rho
    \left[
    \overrightarrow{\prod_{j=1}^L}e^{is_j\sigma_j(\theta_j-\epsilon\Delta\theta_j)}
    \right]
    \nonumber\\
    =&
    \,O(\epsilon^2) .
\end{align}

Next, we consider the case where a single error term in Eq.~(\ref{eq::no_para_gg'}) occurs on one of the inserted Pauli operators $\sigma_{\vec{v}_{j'+1:j'}}$. This leads to Eq.~(\ref{eq::no_para_gg'}) being expressed as
\begin{align}
    &\left(
    \frac{1}{4^n}
    \right)^L\sum_{\vec{v}_1,\ldots,\vec{v}_L}
    \left[
    \sigma_{\vec{v}_L}
    [\cdots ]
    \sigma_{\vec{v}_{j'+2,j'+1}}
    e^{-is_{j'+1}\sigma_{j'+1}(\theta_{j'+1}-\epsilon \Delta\theta_{j'+1})}
    (\sigma_{\vec{v}_{j'+1:j'}}i\epsilon \widetilde{C}_{\vec{v}_{j'+1:j'}})
    e^{-is_{j'}\sigma_{j'}(\theta_{j'}-\epsilon \Delta\theta_{j'})}
    \sigma_{\vec{v}_{j':j'-1}}
    [\cdots]
    \sigma_{\vec{v}_1}\right]
    \rho 
    [\cdots]
    \nonumber\\
    =&
    \left(
    \frac{1}{4^n}
    \right)^L\sum_{\vec{v}_1,\ldots,\vec{v}_L}
    \left[
    \sigma_{\vec{v}_L}
    [\cdots ]
    \sigma_{\vec{v}_{j'+2}}\sigma_{\vec{v}_{j'+1}}
    e^{-is_{j'+1}\sigma_{j'+1}(\theta_{j'+1}-\epsilon \Delta\theta_{j'+1})}
    (\sigma_{\vec{v}_{j'+1}}\sigma_{\vec{v}_{j'}}i\epsilon \widetilde{C}_{\vec{v}_{j'+1:j'}})
    e^{-is_{j'}\sigma_{j'}(\theta_{j'}-\epsilon \Delta\theta_{j'})}
    \sigma_{\vec{v}_{j'}}
    \sigma_{\vec{v}_{j'-1}}
    [\cdots]
    \sigma_{\vec{v}_1}\right]
    \rho 
    [\cdots]
    \nonumber\\
    =&
    \left(
    \frac{1}{4^n}\right)^2
    \sum_{\vec{v}_{j'+1},\vec{v}_{j'}}
    \left(
    \prod_{j=j'+1}^L
    e^{-i\sigma_j(\theta_j-\epsilon\Delta\theta_j)}\right)
    \sigma_{\vec{v}_{j'}}
    i\epsilon \widetilde{C}_{\vec{v}_{j'+1:j'}}
    \sigma_{\vec{v}_{j'}}
    \left(
    \prod_{j=1}^{j'}
    e^{-i\sigma_j(\theta_j-\epsilon\Delta\theta_j)}
    \right)\rho[\cdots]
    \nonumber\\
    =&
    \left(
    \prod_{j=j'+1}^L
    e^{-i\sigma_j(\theta_j-\epsilon\Delta\theta_j)}\right)
    \left\{\frac{1}{4^n}
    \sum_{\vec{v}_{j'}}
    \sigma_{\vec{v}_{j'}}
    \left(
    \frac{1}{4^n}
    \sum_{\vec{v}_{j'+1}}
    \widetilde{C}_{\vec{v}_{j'+1:j'}}
    \right)
    \sigma_{\vec{v}_{j'}}
    \right\}
    \left(
    \prod_{j=1}^{j'}
    e^{-i\sigma_j(\theta_j-\epsilon\Delta\theta_j)}
    \right)\rho[\cdots],
\end{align}
which is also $O(\epsilon^2)$ since
$\sum_{\vec{v}_{j'+1}}\widetilde{C}_{\vec{v}_{j'+1:j'}}=\sum_{\vec{u}}\widetilde{C}_{\vec{u}}$ is independent of $\vec{v}_{j'}$ and has only $O(\epsilon^2)$ term proportional to $I$.

Finally, let us consider how many first- and second-order $\epsilon$ terms can possibly appear in Eq.~(\ref{eq::no_para_gg'}). 
The first-order $\epsilon$ terms can only appear in the expansion of Eq.~(\ref{eq::no_para_gg'}) when a single error term is chosen at either $\sigma_{\vec{v}_1}$ or $\sigma_{\vec{v}_L}$. Thus, the first-order $\epsilon$ terms in Eq.~(\ref{eq::no_para_gg'}) do not scale in terms of $L$ and behave like $O(\epsilon)$. The second-order $\epsilon$ terms can originate from any term in the expansion of Eq.~(\ref{eq::no_para_gg'}) where a single error term occurs at a single Pauli rotation in the computation $e^{\pm is_j\sigma_j(\theta_j-\epsilon\Delta\theta_j)}$ or at an inserted Pauli operator $\sigma_{\vec{v}_{j:j+1}}$; the sum of such terms is $O(\epsilon^2 L)$. Technically speaking, a second-order $\epsilon$ term can also possibly appear when two error terms occur in Eq.~(\ref{eq::no_para_gg'}), but for such cases to have a second-order $\epsilon$ term, the errors must appear in close locations; this means that such cases are largely irrelevant in practice. For example, when the error occurs at $e^{\pm is_j\sigma_j(\theta_j-\epsilon\Delta\theta_j)}$ and $\sigma_{\vec{v}_{k:k+1}}$ with either $j-k\geq 2$ or $k-j\geq 1$, it can be easily shown that the error terms are averaged over random Pauli operators and therefore no more second-order $\epsilon^2$ terms remains. Thus, the total sum of these terms in the expansion of Eq.~(\ref{eq::no_para_gg'}) will also be $O(\epsilon^2 L)$. 

In general, $\epsilon^n$ terms in Eq.~(\ref{eq::no_para_gg'}) can be shown to be $O(\epsilon^n L^{\lfloor \tfrac{n}{2} \rfloor})$, where $\lfloor a\rfloor$ denotes the largest integer not exceeding $a$. This follows from the fact that the first-order $\epsilon$ term in the error can remain only if there exists another error term in a nearby gate. Using this fact, the claim made in Eq.~(\ref{eq::no_para_wanttoshow}) is shown. \hfill \qed

%%%%%%%%%%%%%%%%%%%%%%%%%%%%%%%%%%%%%%%%%%%%%%%%%%%%%%%%%%%%%%%%%%%%%%%%%

\section{Robust Phase Estimation}\label{app::robustphaseestimation}

For the sake of completeness, we reproduce the key ingredients and main theorem of the robust phase estimation technique, developed in Ref.~\cite{kimmel2015robust}.

First, note that any single-qubit unitary gate can be characterized by the following gate set: $Z_{\pi/2}$, a $\tfrac{\pi}{2}$ rotation about the $Z$-axis of the Bloch sphere, and $X_{\pi/4}$, a $\tfrac{\pi}{4}$ rotation about the $X$-axis. Supposing that both of these gates suffer from coherent errors leads to the following general form for their noisy versions:
\begin{align}
    Z_{\pi/2}(\omega) &= \cos{\left(\frac{\pi}{4}\left( 1+\omega\right)\right)} \mathbb{I} - i \sin{\left(\frac{\pi}{4}\left( 1+\omega\right)\right)} \sigma_z, \notag \\
    X_{\pi/4}(\xi,\varphi) &= \cos{\left(\frac{\pi}{8}\left( 1+\xi\right)\right)} \mathbb{I} - i \sin{\left(\frac{\pi}{4}\left( 1+\xi\right)\right)} (\cos{(\varphi)} \sigma_x + \sin{(\varphi)} \sigma_z).
\end{align}
Here, $\omega$ quantifies how far the implemented angle of rotation around the $Z$-axis is from $\tfrac{\pi}{2}$; as the axis of rotation of the second gate lies in the $XZ$-plane, it has two degrees of freedom: $\varphi$ is the angle of the axis of rotation relative to the $X$-axis and $\xi$ quantifies how far the implemented angle of rotation is from $\tfrac{\pi}{4}$.

The phase estimation protocol proceeds as follows to estimate the noise parameters $\omega$, $\xi$, and $\varphi$. Firstly, assuming perfect state preparation of $\ket{0}$, $\ket{+} := \tfrac{1}{\sqrt{2}}(\ket{0}+\ket{1})$, and $\ket{\rightarrow} := \tfrac{1}{\sqrt{2}}(\ket{0}+i\ket{1})$, it is straightforward to verify that
\begin{align}
    |\braket{+|Z_{\pi/2}(\omega)^k|+}|^2 &= \frac{1}{2} \left[ 1+ \cos{\left( -k \frac{\pi}{2} (1+\omega)\right)}\right], \notag \\
    \braket{+|Z_{\pi/2}(\omega)^k|\!\rightarrow}|^2 &= \frac{1}{2} \left[ 1+ \sin{\left( -k \frac{\pi}{2} (1+\omega)\right)}\right],
\end{align}
from which the value of $\omega$ can be determined. Similarly, the value of $\xi$ can be deduced from
\begin{align}
    |\braket{0|X_{\pi/4}(\xi,\varphi)^k|0}|^2 &= \frac{1}{2} \left[ 1+ \cos{\left( k \frac{\pi}{4} (1+\xi)\right)}\right] + \sin^2{\left( k \frac{\pi}{8} (1+\xi) \right)} \sin^2{(\varphi)}, \notag \\
    \braket{0|X_{\pi/4}(\xi,\varphi)^k|\!\rightarrow}|^2 &= \frac{1}{2} \left[ 1+ \sin{\left( k \frac{\pi}{4} (1+\xi)\right)}\right] - \sin{\left( k \frac{\pi}{4} (1+\xi) \right)} \sin^2{(\frac{\varphi}{2})}.
\end{align}
Lastly, by applying the rotation sequence
\begin{align}
    Z_{\pi/2}(0) X_{\pi/4}(\xi,\varphi)^4 Z_{\pi/2}(0)^2 X_{\pi/4}(\xi,\varphi)^4 Z_{\pi/2}(0),
\end{align}
one can estimate $\varphi$ (see Sec. III of Ref.~\cite{kimmel2015robust} for details).

Although above we have assumed the ability to perform perfect state preparations and measurements, this assumption is not necessary; the following theorem highlights how the above methods can be applied to extract noise parameters from observable estimates in general, and the scaling that applies.
\begin{Theorem}[Robust Phase Estimation~\cite{kimmel2015robust}]
    Suppose that one can perform two families of experiments, $\ket{0}$-experiments and $\ket{+}$-experiments, indexed by $k \in \mathbb{Z}^+$, whose probabilities of success are, respectively,
    \begin{align}
        p_0(A, k) &= \frac{1}{2} \left( 1 + \cos{(kA)} \right) + \delta_0(k),\\
        p_+(A, k) &= \frac{1}{2} \left( 1 + \sin{(kA)} \right) + \delta_+(k).
    \end{align}
    Also assume that performing either of the $k^{\textup{th}}$ experiments takes time proportional to $k$, and that
    \begin{align}
        \sup_k \{ |\delta_0(k)|,|\delta_+(k)|\} < 1/\sqrt{8}.
    \end{align}
    Then an estimate $\hat{A}$ of $A\in (-\pi,\pi]$ with standard deviation $\sigma(\hat{A})$ can be obtained in time $T = O(1/\sigma(\hat{A}))$ using non-adaptive experiments.
    On the other hand, if $|\delta_0(k)|$ and $|\delta_+(k)|$ are less than $1/\sqrt{8}$ for all $k<k^*$, then it is possible to obtain an estimate $\hat{A}$ of $A$ with $\sigma(\hat{A}) \sim O(1/k^*)$ (with no promise on the scaling of the procedure).
\end{Theorem}

%%%%%%%%%%%%%%%%%%%%%%%%%%%%%%%%%%%%%%%%%%%%%%%%%%%%%%%%%%%%%%%%%%%

\section{Root Mean Squared Error Bounds}\label{app::diamondnorm}

An upper bound of the RMS of the error of expectation values in terms of an arbitrary observable is as follows.
\begin{lem}\label{le::error_vari}
    For an arbitrary unitary operation defined by $\mathcal{U}(\rho ):=U\rho U^{\dagger}$ with a unitary operator $U$ and a density operator $\rho$ on a Hilbert space $\mathcal{H}$, if a set of deterministic quantum operations (completely-positive trace-preserving maps) $\mathcal{F}_j:\mathcal{L}(\mathcal{H})\to \mathcal{L}(\mathcal{H})$ and a probability distribution $\{ p_j \}$ satisfies
    \begin{align}
        \| \mathcal{U}-\sum_j p_j \mathcal{F}_j \|_{\diamond}\leq \Delta
        \nonumber
    \end{align}
    for some $\Delta >0$, the variance of the averaged operation $\sum_j p_j \mathcal{F}_j$ is upper bounded by 
    \begin{align}
        \underset{\substack{ \mathrm{dim}(\mathcal{H}^{\prime}) \\ \ket{\psi};\|\ket{\psi}\|=1 }}{\mathrm{sup}}
        \sum_j p_j\|\mathcal{U}\otimes \mathcal{I}(\ket{\psi}\bra{\psi})-
        \mathcal{F}_j\otimes \mathcal{I}(\ket{\psi}\bra{\psi}) \| _1 ^2
        \leq 2\Delta,
    \end{align}
    where $\mathcal{I}$ is the identity operation on a Hilbert space $\mathcal{H}^{\prime}$ 
    and $\ket{\psi}$ is a pure state on $\mathcal{H}\otimes \mathcal{H}^{\prime}$.
\end{lem}

The proof is given in App. B of Ref.~\cite{odake2023higher}. For our case, $U$ is the error-free unitary that one desires to implement, $j:=(\vec{v}_1,\ldots,\vec{v}_B)$ is the tuple of all random vectors chosen in the REAS scheme, and $p_j$ and $\mcF_j$ are the probability that $j$ and the quantum operation on $\mcH$ which is applied on the state (including the final partial trace of $\mcH_{\rm env}$) when $j$ is chosen, respectively. 

Consider $\mcE_{\rm shifted}$ to correspond to a gate implemented with a shift in the rotation angle, i.e., the r.h.s.\ of Eq.~\eqref{eq::desc_of_layered-2} of the main text. Since the result regarding such coherent errors implies that $\|\mcE_{\rm shifted}-\mcE'\|_{\diamond}=C(\epsilon+B\epsilon^2)$ for constant $C>0$ and $\mcE'=\sum_j p_j\mcF_j$, where $p_j:=(1/4)^{nB}$ and $\mcF_j(\rho):={\rm tr}_{\rm env}[\mcE'_{j}(\rho\otimes \rho_{\rm env})]$, Lem. \ref{le::error_vari} leads to the following.
\begin{corollary} 
For an arbitrary observable $A$ on $\mcH$ such that $\|A\|_{\rm op}\leq 1$, the following equation holds:
\begin{align}
    \underset{\substack{ \ket{\psi};\|\ket{\psi}\|=1 }}{\mathrm{sup}}
        \sum_j p_j
        [{\rm tr}\{A\mcE_{\rm shifted}(\ket{\psi}\bra{\psi})\}-
        {\rm tr}\{A\mathcal{F}_j(\ket{\psi}\bra{\psi})\} ]^2
        \leq 2C(\epsilon+N\epsilon^2)
\end{align}
where $j:=(\vec{v}_1,\ldots ,\vec{v}_B)$, $p_j=(1/4)^{nB}$, and $\mcF_j(\rho):={\rm tr}_{\rm env}[\mcE'_{j}(\rho\otimes \rho_{\rm env})]$.
\end{corollary}
\noindent This means that the RMS of the difference between the error of expectation value is bounded above by $\sqrt{2C(\epsilon+B\epsilon^2)}$ which behaves as $\sqrt{2C}\epsilon\sqrt{B}$ in the regime $B\gg 1/\epsilon$.

%%%%%%%%%%%%%%%%%%%%%%%%%%%%%%%%%%%%%%%%%%%%%%%%%%%%%%

\section{Relation to Randomized Compiling}\label{app::reasvsrc}

Here, we analyze the relation between REAS and \emph{randomized compiling} (\textbf{RC})~\cite{wallman2016noise}. Although both methods make use of inserting random gates to suppress errors, they do so in different ways, leading to distinct scaling behaviors. In particular, here we present an example such that the following holds for a set of circuits of depth $L$:
\begin{align}
    \|\mcE_{\rm ideal}-\mcE_{\mathrm{RC}}\|_{\diamond }&\geq \beta \epsilon L + O(\epsilon^2) \label{eq::rcbound} \\
    \|\mcE_{\rm ideal}-\mcE_{\mathrm{REAS}}\|_{\diamond }&\leq \alpha \epsilon + O(\epsilon^2).\label{eq::reasbound}
\end{align}
where $\alpha,\beta >0$ are constants, $\epsilon >0$ is the noise parameter specifying the maximum magnitude of the error, $\| \cdot \|_{\diamond}$ denotes the diamond norm, and $\mcE_{\rm REAS}$ and $\mcE_{\rm RC}$ are the quantum channels obtained by applying the REAS and RC schemes, respectively, to the circuit of interest, which is represented by $\mcE_{\rm ideal}$ in the ideal case. Although the error bound for RC may be improvable by optimizing the details of the compilation (e.g., the choice of easy/twirling/hard gates, etc.), the above bound for REAS holds in general. Thus, the example highlights that in the worst case scaling of the error in the limit $\epsilon\to 0$, REAS outperforms RC for this particular circuit. 

We demonstrate the bounds in Eqs.~\eqref{eq::rcbound} and~\eqref{eq::reasbound} for a single-qubit circuit. We begin with the randomized compiling bound~\eqref{eq::rcbound}. Here, in line with the setting of Ref.~\cite{wallman2016noise}, we choose the set of easy gates $\mathbf{C} := \{I,X,Y,Z,S,SX,SY,SZ\}$ as the group generated by the Pauli operators and the phase gate $S$, and the set of hard gates $\mathbf{G} := \{H, T\}$ (control-$Z$ is not included here since we consider a single qubit circuit). 

Consider now the circuit in which all easy and hard gates are chosen as $S$ (phase gate) and $T$ ($\pi /8$-gate), respectively (note that $S=e^{-i(\pi /4)Z}$ and $T=e^{-i(\pi /8)Z}$ up to global phases). Randomized compiling works by inserting particular \emph{twirling gates} at appropriate points throughout the circuit in order to render coherent errors into the form of stochastic Pauli errors. Here, we choose the set of twirling gates $\mathbf{T} := \{I,X,Y,Z\}$ as the Pauli group. We assume that any error is coherent and only occurs on the Pauli $Z$ gate in the form $Z\to Z(I+i\epsilon Z +O(\epsilon^2))$; all other gates in $\mathbf{C}$ and $\mathbf{G}$ have no error. More precisely, defining the error function $g(U)$ of a unitary gate $U$ such that any noisy unitary gate is written as $U[I+i\epsilon g(U) +O(\epsilon^2)]$ and ${\rm tr}[g(U)]=0$ without loss of generality, we have that
\begin{align}\label{eq:errorfunction}
    g(U)=
    \begin{cases}
        Z&(U=Z)\\
        0&(U\in (\mathbf{C}\cup \mathbf{G})\backslash \{Z\})
    \end{cases}.
\end{align}

\noindent The circuit after applying the RC technique transforms as follows:
\begin{figure}[H]
    \centering
    \includegraphics[width=0.8\linewidth]{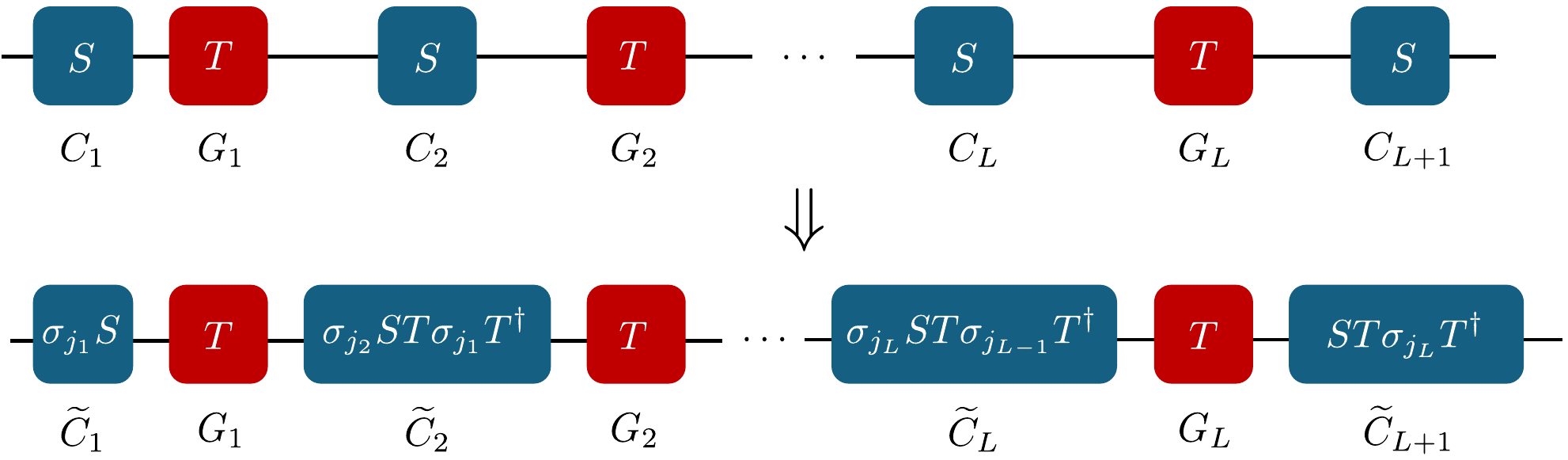}
\end{figure}
\noindent where $\sigma_{j_1},\sigma_{j_2},\ldots ,\sigma_{j_L}$ ($j_1,\ldots,j_L\in \{0,1,2,3\}$) are uniformly randomly sampled Pauli operators.  In the ideal case, we have the transformation $\mcE_{\rm ideal}(\rho):=S^{L+1}T^{L}\rho (T^L)^{\dagger}(S^{L+1})^{\dagger}$. On average, the compiled circuit acts as $\mcE_{\rm RC}(\rho):=\mathbb{E}_{j_1,\ldots ,j_L}[U_{j_1,\ldots,j_L}\rho U_{j_1,\ldots,j_L}^{\dagger}]${, where $U$ appropriately incorporates the easy, twirling, and hard gates, as well as the coherent errors} (see the figure above). Denoting the first-order $\epsilon$ term of $U_{j_1,\ldots ,j_L}$ as $iA_{j_1,\ldots,j_L}$ and defining the expectation $A:=\mathbb{E}_{j_1,\ldots ,j_L}[A_{j_1,\ldots,j_L}]$, we can then write the first-order $\epsilon$ term of $\mcE_{\rm ideal}-\mcE_{\rm RC}$ as $\rho \mapsto [iA\rho (T^L)^{\dagger}-iT^L\rho A^{\dagger}]$. Using the error function $g$ defined in Eq.~\eqref{eq:errorfunction}, the operator $A$ can be written as 
\begin{align}\label{eq::A1}
    A=\mathbb{E}_{j_1,\ldots ,j_L}&
    \left[
    \tilde{C}_{L+1}
    \left({\prod}_{k=1}^LG_k\tilde{C}_k\right)g(\tilde{C}_1)+\tilde{C}_{L+1}g(\tilde{C}_{L+1})
    \left({\prod}_{k=1}^LG_k\tilde{C}_k\right)
    \right.\nonumber\\
    &\quad +\left.
    \sum_{\ell=2}^L
    \tilde{C}_{L+1}
    \left({\prod}_{k=\ell}^{L}G_k\tilde{C}_k\right)g(\tilde{C}_{\ell})
    \left({\prod}_{k=1}^{\ell-1}G_k\tilde{C}_k\right)
    \right] ,
\end{align}
where $G_k$ represent the hard gates and $\tilde{C}_k$ represent the easy gates merged with the adjacent twirl gate. Now note that for $\ell \in \{1,\ldots,L\}$, we have for the considered circuit in particular
\begin{align}
    \prod_{k=1}^{\ell} G_k\tilde{C}_k&=T\sigma_{j_{\ell}}S^{\ell}T^{\ell-1}
    \nonumber\\
    \tilde{C}_{L+1}
    \left(\prod_{k=\ell}^LG_k\tilde{C}_k\right)
    &=
    \begin{cases}
        S^{L+1}T^L&\quad\quad{\rm for}\; \ell=1\\
        S^{L+2-\ell}T^{L+2-\ell} \sigma_{j_{\ell-1}}T^{\dagger}&\quad\quad{\rm otherwise.}
    \end{cases}
\end{align}
With this, Eq.~\eqref{eq::A1} can be rewritten as
\begin{align}\label{eq::A2}
    A&=\mathbb{E}_{j_1,\ldots ,j_L}\left[
    S^{L+1}T^Lg(\sigma_{j_1})+
    ST\sigma_{j_L}T^{\dagger}
    g(ST\sigma_{j_L}T^{\dagger})
    T\sigma_{j_{L}}S^{L}T^{L-1}
    \right.
    \nonumber\\
    &\quad\quad\quad\quad\quad\left.
    +\sum_{\ell=2}^L
    S^{L+2-\ell}T^{L+2-\ell}\sigma_{j_{\ell-1}}T^{\dagger}
    g(\sigma_{j_{\ell}}ST\sigma_{j_{\ell-1}}T^{\dagger})T\sigma_{j_{\ell-1}}S^{\ell-1}T^{\ell-2}
    \right] \nonumber\\
    &=\frac{1}{4}S^{L+1}T^LZ+\sum_{\ell=2}^L
    \mathbb{E}_{j_{\ell}, j_{\ell-1}}
    \left[
    S^{L+2-\ell}T^{L+2-\ell}\sigma_{j_{\ell-1}}T^{\dagger}
    g(\sigma_{j_{\ell}}ST\sigma_{j_{\ell-1}}T^{\dagger})T\sigma_{j_{\ell-1}}S^{\ell-1}T^{\ell-2}
    \right] \nonumber\\
    &=\frac{1}{4}S^{L+1}T^LZ+\sum_{\ell=2}^L
    S^{L+2-\ell}T^{L+2-\ell}
    \mathbb{E}_{j_{\ell-1}}
    \left[\sigma_{j_{\ell-1}}T^{\dagger}
    \mathbb{E}_{j_{\ell}}\{
    g(\sigma_{j_{\ell}}ST\sigma_{j_{\ell-1}}T^{\dagger})\}T\sigma_{j_{\ell-1}}\right]S^{\ell-1}T^{\ell-2}
    .
\end{align}
Now note that $\mathbb{E}_{j_{\ell}}\{
    g(\sigma_{j_{\ell}}ST\sigma_{j_{\ell-1}}T^{\dagger})\}$ depends on $\sigma_{j_{\ell-1}}$ as
\begin{align}\label{eq::merge_depend}
    \mathbb{E}_{j_{\ell}}\{
    g(\sigma_{j_{\ell}}ST\sigma_{j_{\ell-1}}T^{\dagger})\}=
    \begin{cases}
        0& \quad\quad\sigma_{j_{\ell-1}}\in \{I,Z\}\\
        \frac{1}{4}Z&\quad\quad\sigma_{j_{\ell-1}}\in \{X,Y\}.
    \end{cases}
\end{align}
With this, Eq.~(\ref{eq::A2}) becomes
\begin{align}
    A&=\frac{1}{4}S^{L+1}T^L Z+\sum_{\ell=2}^L
    \frac{1}{4}S^{L+2-\ell}T^{L+2-\ell}\left[
    XT^{\dagger}\left(\frac{1}{4}Z\right)TX+
    YT^{\dagger}\left(\frac{1}{4}Z\right)TY
    \right]S^{\ell-1}T^{\ell-2}
    \nonumber\\
    &=\frac{1}{4}S^{L+1}T^L Z-\frac{1}{8}(L-1)S^{L+1}T^LZ.
\end{align}
Therefore, the first-order $\epsilon$ term of $\|\mcE_{\rm ideal}-\mcE_{\rm RC}\|_{\diamond}$ scales linearly in $L$, as claimed in Eq.~\eqref{eq::rcbound}.

Now, we move on to show the validity of Eq.~(\ref{eq::reasbound}). Using the REAS scheme, the circuit transforms as follows: 
\begin{figure}[H]
    \centering
    \includegraphics[width=0.8\linewidth]{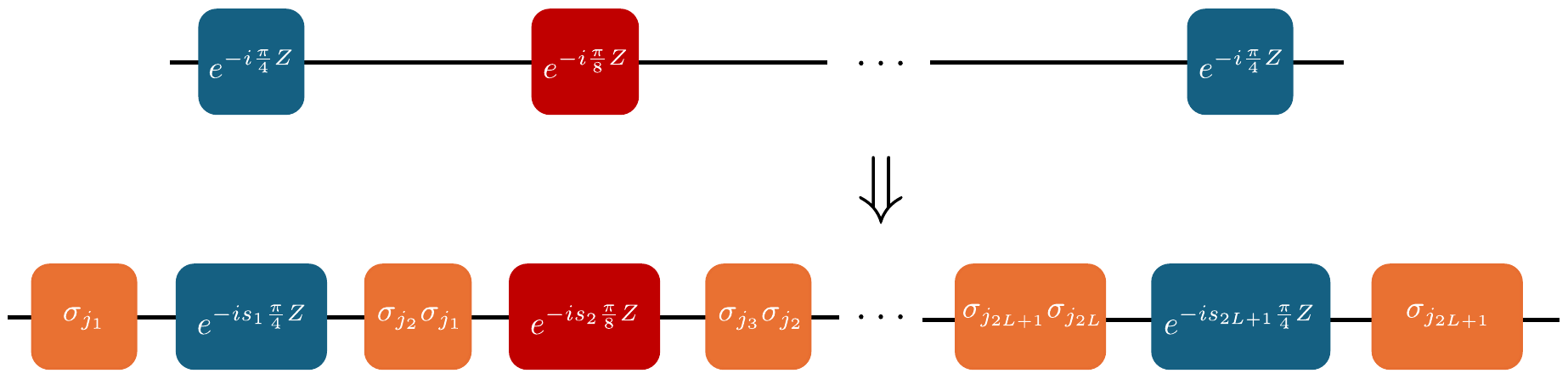}
\end{figure}
\noindent where $\sigma_{j_1},\ldots ,\sigma_{j_{2L+1}}$ are uniformly randomly sampled Pauli operators and $s_k$ is equal to 1 if $\sigma_{j_{k}}$ commutes with $Z$ and set to the value $-1$ otherwise. Note that, in contrast to the above discussion, here we explicitly write $S=e^{-i(\pi /4)Z}$ and $T=e^{-i(\pi /8)Z}$ [which holds up to global phases] since the REAS scheme introduces an $s_k$ factor into the exponent depending upon whether neighboring operations commute or anti-commute. In REAS, the twirling gate set is chosen to be the set of Pauli operators. Since the basic principle is sandwiching the errors on $e^{-is_k\theta_k Z}$ ($\theta_k=\pi/4$ if $k$ is odd and otherwise $\theta_k=\pi/8$) by random Pauli operators, the gates $e^{-is_k\theta_k Z}$ and $\sigma_{j_{k+1}}\sigma_{j_k}$ are the counterparts of the hard gates $G_k$ and the merged gates $\tilde{C}_k$ in RC, respectively. In REAS, hard gates are transformed as $e^{-i\theta_k Z}\mapsto e^{-is_k\theta_k Z}$ to ensure that $e^{-i\theta_kZ}=\sigma_{j_k}e^{-is_k\theta_kZ}\sigma_{j_k}$, which is different to how the twirling gates $\sigma_{j_k}$ transform the circuit in the RC method. Thus, the REAS approach requires additional necessity to deal with errors on gates of the form $e^{\pm i \sigma_k \theta_k}$, which can depend on $s_k$ and are therefore not canceled by averaging over Pauli operators (as they are in RC). The aforementioned calibration and single Pauli transformation subroutines of the main text can deal with such errors (but are based upon a particular model of decoherence error; see App.~\ref{app::decoherence}). Of course, such dependence, together with the fact that REAS can increase the depth of the circuit (albeit only to at most twice as long), can be thought of as a disadvantage of REAS compared to RC. 

On the other hand, these differences makes REAS more robust against gate-dependent errors on the twirling gates. Defining $A:=\mathbb{E}_{j_1,\ldots ,j_{2L+1}}[A_{j_1,\ldots,j_{2L+1}}]$, where $iA_{j_1,\ldots,j_{2L+1}}$ is the first-order $\epsilon$ term of the unitary operator obtained by applying REAS to the circuit with random variables chosen to be $(j_1,\ldots,j_{2L+1})$, we have
\begin{align}\label{eq::reas_ss0}
    A&=\mathbb{E}_{j_1,\ldots ,j_{2L+1}}
    \left[S^{L+1}T^{L}g(\sigma_{j_1})+\sigma_{j_{2L+1}}g(\sigma_{j_{2L+1}})\sigma_{j_{2L+1}}S^{L+1}T^{L}
    \right.
    \nonumber\\
    &\quad\quad\quad\quad\quad\left.
    +\sum_{\ell=2}^{2L+1}\left(\prod_{k=\ell}^{2L+1}e^{-i\theta_{j_k}Z}\right)\mathbb{E}_{j_{\ell}, j_{\ell-1}}[\sigma_{j_{\ell-1}}g(\sigma_{j_{\ell}}\sigma_{j_{\ell-1}})\sigma_{j_{\ell-1}}]\left(\prod_{k=1}^{\ell-1}e^{-i\theta_{j_k}Z}\right)
    \right]
    \nonumber\\
    &= O(1) +\sum_{\ell=2}^{2L+1}\left(\prod_{k=\ell}^{2L+1}e^{-i\theta_{j_k}Z}\right)\mathbb{E}_{ j_{\ell-1}}[\sigma_{j_{\ell-1}}\mathbb{E}_{j_{\ell}}\{g(\sigma_{j_{\ell}}\sigma_{j_{\ell-1}})\}\sigma_{j_{\ell-1}}]\left(\prod_{k=1}^{\ell-1}e^{-i\theta_{j_k}Z}\right)
    .
\end{align}
Unlike in the case of RC [see Eq.~(\ref{eq::merge_depend})], here $\mathbb{E}_{j_{\ell}}[g(\sigma_{j_{\ell}}\sigma_{j_{\ell-1}})]$ is always equal to $(1/4)[g(I)+g(X)+g(Y)+g(Z)]$ and is thus independent of $j_{\ell-1}$. This property makes the second term of r.h.s.\ of Eq.~(\ref{eq::reas_ss0}) vanish, thereby proving Eq.~(\ref{eq::reasbound}). \qed

The error analysis for REAS shown above applies to general circuits, and as a consequence, the following inequality for the root mean square of the single-shot error can be shown (see App.~\ref{app::diamondnorm}):
\begin{align}
    \sup_{\substack{{\rm dim}\mcH'\\\ket{\psi};\|\ket{\psi}\|=1}}\mathbb{E}_{j}
    \|(\mcE_{\rm ideal}\otimes \mcI)(\ket{\psi}\bra{\psi})-(\mcE_{{\rm REAS}, j}\otimes \mcI)(\ket{\psi}\bra{\psi})\|_1^2\leq 2C(\epsilon +L\epsilon^2), 
\end{align}
where $C$ is a constant, $\mcI$ is the identity map on $\mcH'$, $\ket{\psi}$ is a pure state on $\mcH\otimes \mcH'$, and $\mcE_{{\rm REAS}, j}$ is the single-shot REAS circuit when random variable $j$ is chosen. This inequality shows that the error scales as $\epsilon \sqrt{L}$ in the regime $L\gg 1/\epsilon$. 

\begin{figure}[h]
    \centering
    \includegraphics[width=0.4\linewidth]{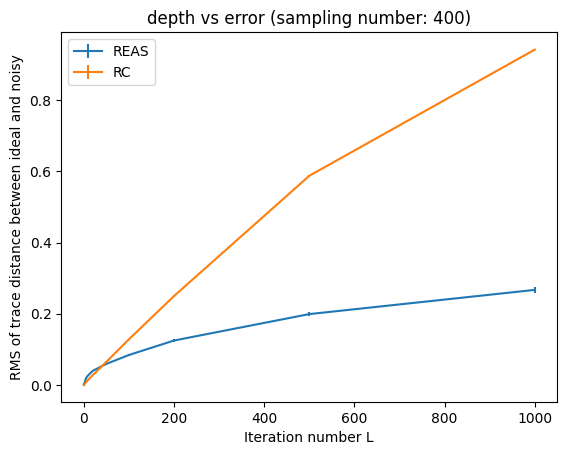}
    \caption{RMS error for both REAS and RC methods measured by the trace distance for a fixed input state. \label{fig::app-reasvsrc}
    }
\end{figure}

To highlight this, in Fig.~\ref{fig::app-reasvsrc} we present the following numerical simulation of the example considered, which shows that REAS indeed achieves quadratic error suppression and outperforms RC in this case. Nonetheless, we note also that for some classes of circuits, numerical simulation shows that RC can also achieve a quadratic error suppression comparable to REAS. Thus, for particular circuits, the error scaling between RC and REAS behaves completely differently. In summary, there is no strict hierarchy between the efficacy of the methods; rather, they are each suitable to a different type of error suppression.

\FloatBarrier

%%%%%%%%%%%%%%%%%%%%%%%%%%%%%%%%%%%%%%%%%%%%%%%

\section{Class of Decoherence Errors that can be Suppressed}\label{app::decoherence}

We now move to demonstrate the more general class of decoherence errors that can be suppressed when our REAS method is supplemented by the single Pauli transformation introduced in Sec.~\ref{subsec::decoherence} of the main text. Specifically, we show that this protocol can suppress errors that are first-order in $\epsilon$ for a particular class of open system-environment dynamics, which includes depolarizing, amplitude-damping, and decoherence models. We assume that the interaction Hamiltonian involves at most two-body interactions.

In attempting to extend the error-suppressing method for the case of coherent errors to the open setting of decoherence, one immediately runs into problem that some errors terms on the computational Pauli gates $e^{-is_j\sigma_j\theta_j}$ cannot be dealt with. In particular, any term of the form $s_j\sigma_j\otimes M_{\rm env}$, where $M_{\rm env}\neq I_{\rm env} \in \mcH_{\rm env}$, cannot be absorbed into the shift of the rotation angle $\theta_j$, since $\sigma_{j}$ is coupled with $M_{\rm env}$, and therefore cannot be erased by averaging over random inserted Pauli operators as in Eqs.~\eqref{eq::randompauliaverageidentity} and \eqref{eq::no_para_s}. This inability to suppress such terms can introduce unwanted $O(\epsilon L)$ errors in Eq.~(\ref{eq::no_para_wanttoshow}), since such first-order $\epsilon$ terms can appear in all Pauli computation gates throughout the circuit.

We circumvent this issue by modifying the computational gate sequence implementing $e^{-is_j\sigma_j\theta_j}$ in such a way that the error terms proportional to $s_j\sigma_j\otimes M_{\rm env}$ for $M_{\rm env}\neq I$ is at most $O(\epsilon^2)$. This \emph{single Pauli transformation} consists of decomposing any $e^{-is\sigma\theta}$ into (a) a two-qubit Pauli gate dependent on the rotation direction $s\in \{+1,-1\}$ and (b) another Pauli gate (which can be either a one- or two-qubit Pauli gate) that is independent of $s$. Intuitively, the trick works to suppress error terms that are first-order in $\epsilon$ since these only appear when an error occurs on a single gate in the computation $e^{-is_j\sigma_j\theta_j}$ and: (a) the $s$-dependent two-qubit Pauli gate does not produce harmful first-order $\epsilon$ terms; and (b) the other $s$-independent Pauli gate does not produce any error term that depends upon $s$. 

First, we show that any gate $e^{-is\sigma_{\vec{v}_0}\theta}$ corresponding to a two-qubit Pauli operator involves only $O(\epsilon^2)$ terms proportional to $s\sigma_{\vec{v}_0}\otimes M_{\rm env}$ for $M_{\rm env}\neq I$. Note that we always take the Hamiltonian to be traceless without loss of generality (since any trace part only contributes to a global phase).
\begin{lem}\label{le::appb_twoq_noerr}
    Consider a time interval $t\in [0,T]$, any two-qubit Pauli operator $\sigma_{\vec{v}_0}\in \mcL(\mcH)$, a time-dependent interaction Hamiltonian $\epsilon H'(t)\in \mcL(\mcH\otimes \mcH_{\rm env})$ which contains at most two-body terms and has operator norm $O(\epsilon)$, a background Hamiltonian $H_{\rm env}$ of the environment, and a function $\Omega(t)$ such that $\int_0^T{\rm d}\,\tau \Omega(\tau)=\theta$. In this setting, the total Hamiltonian $s\Omega(t)\sigma_{\vec{v}_0}\otimes I_{\rm env}+I\otimes H_{\rm env}+\epsilon H'(t)$ gives rise to the unitary operator
    \begin{align}\label{eq::app-lemma1}
        \mcT\exp\left[
        -i\int_0^t{\rm d}\tau
        (s\Omega(t)\sigma_{\vec{v}_0}\otimes I_{\rm env}+I\otimes H_{\rm env}+\epsilon H'(t))
        \right]
        =:&
        \left(
        e^{-is\sigma_{\vec{v}_0}\theta}\otimes
        e^{-iH_{\rm env}T}
        \right)\left[
        I+i\epsilon C_{\vec{v}_0}
        \right]
        \nonumber\\
        =:&
        \left(
        e^{-is\sigma_{\vec{v}_0}\theta}\otimes
        e^{-iH_{\rm env}T}
        \right)\left[
        I+i\epsilon \sum_{\vec{w}\in \{
        0,1,2,3\}^n}
        \sigma_{\vec{w}}\otimes D_{\vec{v}_0; \vec{w}}(s)
        \right],
    \end{align}
    where $\epsilon D_{\vec{v}_0; \vec{v}_0}(s)=\epsilon \alpha I+O(\epsilon^2)$ for some $\alpha$.
\end{lem}
\noindent Note that the assumption that the operator norm of the interaction Hamiltonian is $O(\epsilon)$ is consistent with the assumption that the error term $C_{\vec{v}_0}$ has operator norm of the same order.\\

\textbf{Proof:}~
For general time-dependent Hamiltonians $H_1(t)$ and $H_2(t)$, the Hamiltonian dynamics generated by $H_1+H_2$ can be written as
\begin{align}
    \mcT\exp\left[
    -i\int_0^t{\rm d}\tau
    \{H_1(\tau)+H_2(\tau)\}
    \right]
    =U_1(t)
    \mcT\exp\left[
    -i\int_0^t
    {\rm d}\tau U_1(\tau)^{\dagger}
    H_2(\tau)U_1(\tau)
    \right],
\end{align}
where
\begin{align}
    U_1(t):=\mcT\exp
    \left[
    -i\int_0^t{\rm d}\tau H_1(\tau)
    \right] .
\end{align}
Substituting $H_1\gets s\Omega \sigma_{\vec{v}_0}+H_{\rm env}$ and $H_2\gets \epsilon H'$ into the above expression and taking the upper limit of the integration up to time $T$ leads to
\begin{align}\label{eq::appb_mcTevolution}
    \mcT\exp&\left[
    -i\int_0^T{\rm d}\tau
    (s\Omega(\tau)\sigma_{\vec{v}_0}\otimes I_{\rm env}+I\otimes H_{\rm env}+\epsilon H'(\tau))
    \right]
    =
    \left(
    e^{-is\theta\sigma_{\vec{v}_0}}
    \otimes
    e^{-iH_{\rm env}T}
    \right)
    \mcT\exp\left[
    -i\epsilon \int_0^T{\rm d}\tau
    U(\tau)^{\dagger}
    H'(\tau)
    U(\tau)
    \right]
    \nonumber\\
    &=\left(
    e^{-is\theta\sigma_{\vec{v}_0}}
    \otimes
    e^{-iH_{\rm env}T}
    \right)
    \left[
    I-i\epsilon \int_0^T{\rm d}\tau
    U(\tau)^{\dagger}
    H'(\tau)
    U(\tau) +O(\epsilon^2)
    \right],
\end{align}
where we have defined
\begin{align}
    U(t):=e^{
    -is\omega(t)
    \sigma_{\vec{v}_0}}
    \otimes e^{-iH_{\rm env}t} \quad\quad\quad \omega(t):=\int_0^t{\rm d}\tau
    \Omega(\tau).
\end{align}
Now, by decomposing the interaction Hamiltonian $H'(t)$ as
\begin{align}
    H'(t):=&\sum_{\vec{w}}
    \sigma_{\vec{w}}\otimes H'_{\vec{w}}(t) ,
\end{align}
Eq.~\eqref{eq::appb_mcTevolution} can be rewritten as
\begin{align}
    &\left(
    e^{-is\theta\sigma_{\vec{v}_0}}
    \otimes
    e^{-iH_{\rm env}T}
    \right)
    \left[
    I-i\epsilon \int_0^T{\rm d}\tau
    U(\tau)^{\dagger}
    \left(
    \sum_{\vec{w}}
    \sigma_{\vec{w}}\otimes H'_{\vec{w}}(t)
    \right)
    U(\tau) +O(\epsilon^2)
    \right]
    \nonumber\\
    =&
    \left(
    e^{-is\theta\sigma_{\vec{v}_0}}
    \otimes
    e^{-iH_{\rm env}T}
    \right)
    \left[
    I-i\epsilon \int_0^T{\rm d}\tau
    \left(
    \sum_{\vec{w}}
    (e^{is\omega(t)\sigma_{\vec{v}_0}}\sigma_{\vec{w}}e^{-is\omega(t)\sigma_{\vec{v}_0}})\otimes (e^{iH_{\rm env}t}H'_{\vec{w}}(t)e^{-iH_{\rm env}t})
    \right)
    +O(\epsilon^2)
    \right].
\end{align}
Lastly, note that $e^{is\omega(t)\sigma_{\vec{v}_0}}\sigma_{\vec{w}}e^{-is\omega(t)\sigma_{\vec{v}_0}}$ contains a $\sigma_{\vec{v}_0}$ component if and only if $\vec{w}=\vec{v}_0$, which can be verified using 
\begin{align}
    {\rm tr}\left[\sigma_{\vec{v}_0}
    (e^{is\omega(t)\sigma_{\vec{v}_0}}\sigma_{\vec{w}}e^{-is\omega(t)\sigma_{\vec{v}_0}})
    \right]
    ={\rm tr}\left[\sigma_{\vec{v}_0}
    \sigma_{\vec{w}}
    \right].
\end{align}
This finally leads us to be able to write $D_{\vec{v}_0;\vec{v}_0}$ [see Eq.~\eqref{eq::app-lemma1}] as
\begin{align}
    D_{\vec{v}_0;\vec{v}_0}=
    -\int_0^T{\rm d}\tau
    (e^{iH_{\rm env}t}H'_{\vec{v}_0}(t)e^{-iH_{\rm env}t})
    +O(\epsilon^2).
\end{align}
Here, $H'_{\vec{v}_0}(t)$ is proportional to the identity only in the case where $\sigma_{\vec{v}_0}$ is a two-qubit Pauli gate. This follows by assumption, since the total term $\sigma_{\vec{v}_0}\otimes H'_{\vec{v}_0}(t)$ being of weight at most two implies that $H'_{\vec{v}_0}(t)$ must be of weight zero; this logic breaks down for single-qubit $\sigma_{\vec{v}_0}$, as will discuss below. Thus $\epsilon D_{\vec{v}_0;\vec{v}_0}=\epsilon\alpha I_{\rm env}+O(\epsilon^2)$ for some $\alpha$, as required.
\qed

From this lemma, one can see that any two-qubit Pauli gate does not involve any first-order $\epsilon$ terms which cannot be dealt with by the REAS method. However, crucially, the same logic does not hold in the case where $\sigma_{\vec{v}_0}$ is a \emph{single-qubit} Pauli operator, which can be seen by considering the following counterexample:
\begin{align}
    e^{-i(sX\otimes I_{\rm env}+\epsilon s X\otimes X_{\rm env})t}=
    (e^{-isXt}\otimes I_{\rm env})e^{-i\epsilon s (X\otimes X_{\rm env}) t}
    =(e^{-isXt}\otimes I_{\rm env})
    [I-i\epsilon s t(X\otimes X_{\rm env}) +O(\epsilon^2)]
\end{align}
which has a first-order $\epsilon$ term proportional to $s(X\otimes X_{\rm env})$. Therefore, we simply modify any single-qubit Pauli gates into two-qubit ones by making use of the single Pauli transformation defined in Eq.~(\ref{eq::singlepaulitransformation}) of the main text. In this way, we can remove any first-order $\epsilon$ terms, as formalized below.

\begin{Theorem}\label{th::spt}
For any interaction Hamiltonian with an operator norm $O(\epsilon)$ that contains at most two-body terms, the gate sequences 
\begin{align}\label{eq::appb_spt}
    &e^{isYIt}\to e^{i\tfrac{\pi}{4}XZ}e^{isZZt}e^{-i\tfrac{\pi}{4}XZ}
    \nonumber\\
    &e^{isZIt}\to e^{-i\tfrac{\pi}{4}XI}e^{i\tfrac{\pi}{4}XZ}e^{isZZt}e^{-i\tfrac{\pi}{4}XZ}e^{i\tfrac{\pi}{4}XI}
    \nonumber\\
    &e^{isXIt}\to e^{i\tfrac{\pi}{4}ZI}e^{i\tfrac{\pi}{4}XZ}e^{isZZt}e^{-i\tfrac{\pi}{4}XZ}e^{-i\tfrac{\pi}{4}ZI}.
\end{align}
implement single qubit Pauli gates $e^{-is\sigma_{\vec{v}_0}}$ without giving rise to any first-order $\epsilon$ error terms of form $s \sigma_{\vec{v}_0}\otimes M_{\rm env}$. More precisely, any noisy gate $e^{-is\sigma_{\vec{v}_0}}$ leads to
\begin{align}
    (e^{-is\sigma_{\vec{v}_0}}\otimes U_{\rm env})
    \left[I+i\epsilon \sum_{\vec{w}}
    \sigma_{\vec{w}}\otimes 
    (\overline{D}'_{\vec{v}_0; \vec{w}} +
    s\Delta D'_{\vec{v}_0; \vec{w}} 
    )
    \right]
\end{align}
for some unitary $U_{\rm env}\in \mcH_{\rm env}$, for which $\epsilon\Delta D'_{\vec{v}_0;\vec{v}_0}=\epsilon\alpha I+O(\epsilon^2)$ for some $\alpha$.
\end{Theorem}
\noindent Note that in Eq.~(\ref{eq::appb_spt}), the only $s$-dependency on the r.h.s.\ occurs through the term $e^{isZZt}$ in the middle. It is straightforward to see that the proof of Thm. \ref{th::spt} can be generalized to the case where any other two-qubit Pauli gate is used to incorporate the $s$-dependency, e.g., $e^{isZXt}$, as long as the overall gate sequences lead to the correct single Pauli rotation gates. \\

\textbf{Proof:}~We explicitly show the proof for only the last equation of Eq.~(\ref{eq::appb_spt}); the logic can be easily generalized to demonstrate the other two. First, note that $e^{\pm i\frac{\pi}{4}\sigma}$ for all Pauli operators $\sigma$ are Clifford operators acting as
\begin{align}
e^{i\frac{\pi}{4}\sigma}\sigma'e^{-i\frac{\pi}{4}\sigma}=
\begin{cases}
    \sigma^{\prime}&\quad\quad\sigma\ \mathrm{and} \ \sigma^{\prime} \ \mathrm{commutes}\\
    \frac{i}{2}[\sigma ,\sigma^{\prime}]&\quad\quad\sigma\ \mathrm{and} \ \sigma^{\prime} \ \textup{anti-commutes}\\
\end{cases}.
\end{align}
Combining this with the identity $Ue^{-iHt}U^{\dagger}=e^{-iUHU^{\dagger}t}$ for unitary $U$ and Hamiltonian $H$, it can be easily checked that
\begin{align}
    e^{isXIt}= e^{i\tfrac{\pi}{4}ZI}e^{i\tfrac{\pi}{4}XZ}e^{isZZt}e^{-i\tfrac{\pi}{4}XZ}e^{-i\tfrac{\pi}{4}ZI}
\end{align}
in the error-free situation. 

Now, suppose that $e^{is(ZZ\otimes I)t}$ (i.e., $I$ is taken to act on the rest of $\mcH$) leads to an $s$-dependent error of the form 
\begin{align}
    e^{is(ZZ\otimes I)t}\to
    (e^{is(ZZ\otimes I)t}\otimes U_{\rm env})
    \left[I+i\epsilon \sum_{\vec{w}}
    \sigma_{\vec{w}}\otimes 
    (\overline{D}_{\vec{w}}
    +s\Delta D_{\vec{w}}
    )
    \right]
\end{align}
for some unitary $U_{\rm env}$. For the $s$-independent gates, the errors induced are of the form
\begin{align}
    G_j\to
    (G_j\otimes U_{{\rm env},j})\left[I+i\epsilon \sum_{\vec{w}}
    \sigma_{\vec{w}}\otimes 
    C_{\vec{w}, j}
    \right]
\end{align}
where $G_1,\ G_2,\ G_3,$ and $G_4$ refers to $e^{i\frac{\pi}{4}(ZI\otimes I)},\ $$e^{i\frac{\pi}{4}(XZ\otimes I)},\ $$e^{-i\frac{\pi}{4}(XZ\otimes I)},\ $ and $e^{-i\frac{\pi}{4}(ZI\otimes I)},\ $respectively. 
Then, the 0$^\textup{th}$-order $\epsilon$ term in the noisy version of the sequence $e^{i\tfrac{\pi}{4}ZI}e^{i\tfrac{\pi}{4}XZ}e^{isZZt}e^{-i\tfrac{\pi}{4}XZ}e^{-i\tfrac{\pi}{4}ZI}$ that implements the desired single Pauli rotation $e^{isXIt}$ can be written as
\begin{align}
    e^{is(XI\otimes I)t}\otimes \widetilde{U},
\end{align}
where 
\begin{align}
    \widetilde{U}:=
    U_{{\rm env},1}U_{{\rm env},2}U_{{\rm env}}
    U_{{\rm env},3}U_{{\rm env},4}.
\end{align}

We now move to analyze the first-order $\epsilon$ error term. Such a term in the noisy version of $e^{i\tfrac{\pi}{4}ZI}e^{i\tfrac{\pi}{4}XZ}e^{isZZt}e^{-i\tfrac{\pi}{4}XZ}e^{-i\tfrac{\pi}{4}ZI}$ appears when an error occurs at a single gate in the sequence. Suppose for now that said error occurs at $e^{is(ZZ\otimes I)t}$; then the term of interest is of the form
\begin{align}
    (G_1\otimes U_{{\rm env},1})(G_2\otimes U_{{\rm env},2})(e^{is(ZZ\otimes I)t}\otimes U_{\rm env})\left[
    i\epsilon \sum_{\vec{w}}
    \sigma_{\vec{w}}\otimes 
    (\overline{D}_{\vec{w}}
    +s\Delta D_{\vec{w}}
    )
    \right](G_3\otimes U_{{\rm env},3})(G_4\otimes U_{{\rm env},4}) .
\end{align}
This expression can be rewritten as
\begin{align}
    &i\epsilon
    \sum_{\vec{w}}
    \left(
    G_1G_2e^{is(ZZ\otimes I)t}G_3G_4\cdot G_4^{\dagger}G_3^{\dagger}
    \sigma_{\vec{w}}
    G_3G_4
    \right)
    \otimes
    (U_{{\rm env},1}U_{{\rm env},2}U_{{\rm env}}
    (\overline{D}_{\vec{w}}
    +s\Delta D_{\vec{w}}
    )
    U_{{\rm env},3}U_{{\rm env},4}
    )
    \nonumber\\
    &=
    i\epsilon
    \sum_{\vec{w}}
    \left(
    e^{is(XI\otimes I)t}\cdot G_1G_2
    \sigma_{\vec{w}}
    G_3G_4
    \right)
    \otimes
    (U_{{\rm env},1}U_{{\rm env},2}U_{{\rm env}}
    (\overline{D}_{\vec{w}}
    +s\Delta D_{\vec{w}}
    )
    U_{{\rm env},3}U_{{\rm env},4}
    )
    \nonumber\\
    &=
    (e^{is(XI\otimes I)t}\otimes \widetilde{U})
    \left[
    i\epsilon
    \sum_{\vec{w}}
    (G_1G_2
    \sigma_{\vec{w}}
    G_3G_4)\otimes
    U_{{\rm env},4}^{\dagger}U_{{\rm env},3}^{\dagger}
    (\overline{D}_{\vec{w}}
    +s\Delta D_{\vec{w}}
    )U_{{\rm env},3}U_{{\rm env},4}
    \right]
    ,
\end{align}
using the fact that $G_3^{\dagger}=G_2$ and $G_4^{\dagger}=G_1$. Since $G_1G_2\sigma_{\vec{w}_0}G_3G_4=XI\otimes I$ for $\sigma_{\vec{w}_0}=(ZZ\otimes I)$, the term with $s(XI\otimes I)$ on the system part in the error term can be expressed as
\begin{align}\label{eq::appb_twoq_s_chosen}
    i\epsilon s(XI\otimes I)
    \otimes 
    (U_{{\rm env},4}^{\dagger}U_{{\rm env},3}^{\dagger}
    \Delta D_{\vec{w}_0}
    U_{{\rm env},3}U_{{\rm env},4}) 
\end{align}
Since $\epsilon\Delta D_{\vec{w}_0}=\epsilon\alpha I+O(\epsilon^2)$ by Lem. \ref{le::appb_twoq_noerr}, Eq.~(\ref{eq::appb_twoq_s_chosen}) does not have the first-order $\epsilon$ term where the environment part is not proportional to the identity.

Next, we consider the case where an error term occurs at one of $G_1,\ldots ,G_4$. We only explicitly show the proof for the case where error term is chosen at $G_1$; the other cases follow similarly. The corresponding term is expressed as 
\begin{align}
    &(G_1\otimes U_{{\rm env},1})
    \left[i\epsilon
    \sum_{\vec{w}}
    \sigma_{\vec{w}}\otimes C_{\vec{w},1}
    \right]
    (G_2\otimes U_{{\rm env},2})(e^{is(ZZ\otimes I)t}\otimes U_{\rm env})(G_3\otimes U_{{\rm env},3})(G_4\otimes U_{{\rm env},4}) 
    \nonumber\\
    &=
    i\epsilon\sum_{\vec{w}}
    (G_1\sigma_{\vec{w}}G_2e^{is(ZZ\otimes I)t}G_3G_4)\otimes
    (U_{{\rm env},1}C_{\vec{w},1}U_{{\rm env},2}U_{{\rm env}}U_{{\rm env},3}U_{{\rm env},4})
    \nonumber\\
    &=
    (e^{is(XI\otimes I)t}\otimes \widetilde{U})
    \left[i\epsilon\sum_{\vec{w}}
    (G_2e^{is(ZZ\otimes I)t}G_3G_4)^{\dagger}
    \sigma_{\vec{w}}
    (G_2e^{is(ZZ\otimes I)t}G_3G_4) 
    \right.
    \nonumber\\
    &\quad\quad\quad\quad\quad\quad\quad\quad\left.
    \otimes 
    (U_{{\rm env},2}U_{{\rm env}}
    U_{{\rm env},3}U_{{\rm env},4})^{\dagger}
    C_{\vec{w},1}
    (U_{{\rm env},2}U_{{\rm env}}
    U_{{\rm env},3}U_{{\rm env},4})
    \right] ,
\end{align}
which displays no $s$-dependence, and thus cannot correspond to any term with $s(XI\otimes I)$ on the system part. 

Therefore, in total, the noisy version of the sequence $G_1G_2e^{isZZt}G_3G_4$ is written as
\begin{align}
    (e^{is(XI\otimes I)t}\otimes \widetilde{U})
    \left[I+i\epsilon \sum_{\vec{w}}
    \sigma_{\vec{w}}\otimes 
    (\overline{D}'_{\vec{w}}+s\Delta D'_{\vec{w}})
    \right]
\end{align}
where $\epsilon \Delta D'_{\vec{w}_0}=\epsilon \alpha I+O(\epsilon^2)$ for $\sigma_{\vec{w}_0}:=XI\otimes I$ and some $\alpha$, as required.
\qed

Finally, combining Lem.~\ref{le::appb_twoq_noerr} and Thm.~\ref{th::spt}, we can show that for decoherence errors on all Pauli rotation gates $e^{-is\sigma_{\vec{v}_0}\theta}$ can be dealt with by implementing the gates in such a way that their noisy version is written as
\begin{align}
    e^{-is\sigma_{\vec{v}_0}\theta}\to
    (e^{-is\sigma_{\vec{v}_0}\theta}\otimes U)
    \left[I+i\epsilon \sum_{\vec{w}}\sigma_{\vec{w}}\otimes
    (\overline{D}'_{\vec{w}}+s\Delta D'_{\vec{w}})
    \right]
\end{align}
for $(\vec{v}_0, \theta)$-dependent unitary $U$ and linear operators $\overline{D}'_{\vec{w}}$ and $\Delta D'_{\vec{w}}$, where $\epsilon\Delta D'_{\vec{v}_0}=\epsilon\alpha I+O(\epsilon^2)$ for some $\alpha$. This means that the first-order $\epsilon$ error term of $e^{-is\sigma_{\vec{v}_0}\theta}$ is classified into the cases
\begin{enumerate}
    \renewcommand{\labelenumi}{(\alph{enumi})}
    \item $I\otimes M$
    \item $\sigma_{\vec{w}}\otimes M$ for $\vec{w}\neq \vec{0}$
    \item $s\sigma_{\vec{w}}\otimes M$ for $\vec{w}\neq \vec{v}_0$
    \item $\alpha s\sigma_{\vec{w}}\otimes I_{\rm env}$, 
\end{enumerate}
where $M$ is any linear operator and $\alpha$ is arbitrary.
Terms of type (a) can be avoided by absorbing them into the definition of $U$ (namely, change $U\to Ue^{i\epsilon M^{(0)}}$ where $M^{(0)}$ is the 0$^\textup{th}$-order $\epsilon$ term of $M$). Terms of types (b) and (c) can be erased by averaging over random Pauli operators in a similar way to coherent errors described in App.~\ref{app::reascalibration}. Finally, terms of the form (d) can be absorbed into a shift in the rotation angle. Therefore, we have demonstrated that our proposed REAS scheme can deal with decoherence errors appearing as an $O(\epsilon)$ interaction Hamiltonian with at most two-body interaction in a similar way to how they deal with coherent errors.

%%%%%%%%%%%%%%%%%%%%%%%%%%%%%%%%%%%%%%%%%%%%%%%%%%%%%%%%%%%%%%%%
\section{Details of Experimental Demonstration}\label{app::experiments}

\subsection{Implementation of $ZZ$ Rotation Gates}

A $ZZ(\theta_i)$ rotation gate is implemented via the following sequence of gates $IH$ -- $ECR(\theta_i)$ -- $XI$ -- $IH$, where $H$ denotes a Hadamard gate and $ECR(\theta_i)$ denotes an \textit{Echoed Cross Resonance} (\textbf{ECR}) gate with rotation angle $\theta_i$, which is determined by the amplitude of the CR pulses. For the CR pulses, we used the default CR pulse shape provided by IBM Quantum systems---namely, square Gaussian (square with Gaussian ramps)---and modified their amplitudes by a factor of ten. Here, we changed only the control pulse amplitude and left the other pulse instructions, which are calibrated for the original ECR angle $\frac{\pi}{2}$, unchanged. This suggests that the resulting pulse sequence can implement a $ECR(\theta_i)$ gate with coherent errors; accumulation of such coherent errors throughout the circuit can be well-suppressed by REAS as discussed in Sec.~\ref{sec::reas} of the main text.

\FloatBarrier

\subsection{Experimental Demonstrations for Individual Observables} 
\begin{figure}[h]
    \centering
    \includegraphics[width=\linewidth]{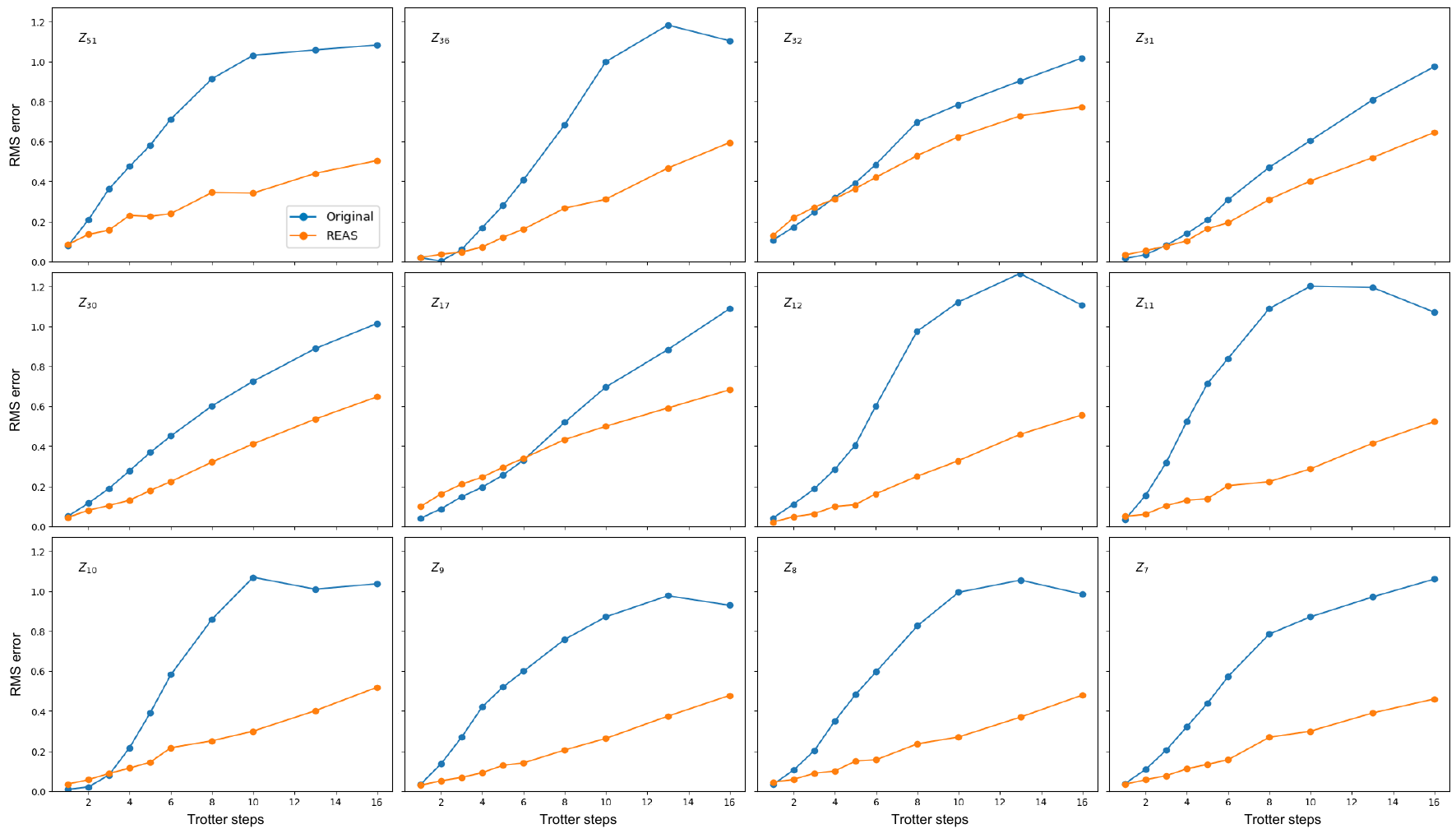}
    \caption{\textit{Experimental Demonstrations for Individual Observables.---}RMS errors of 12 (out of 100) individual weight-1 $Z$ observables by Trotter steps with REAS (orange, REAS) and without REAS (blue, Original) using \texttt{ibm\_brisbane} processor. The label $Z_q$ in the top-left corner of each figure means measuring weight-1 $Z$ observable on physical qubit $q$ of the processor.}
    \label{fig::pauli_z_individual_brisbane}
\end{figure}

In Fig.~\ref{fig::pauli_z_individual_brisbane}, we show the plots of the RMS errors of 12 (out of 100) individual weight-1 $Z$ observables by Trotter steps experimented on \texttt{ibm\_brisbane}. The individual plots show different scaling depending on the observables considered. Figure~\ref{fig::main_exp_result} in the main text shows the average of those plots.

We summarize the qubits used in the experiments and the median device properties on those qubits in Table~\ref{tab:device}.

\begin{table*}[h]
	\centering
	\caption{\textit{Qubits used.---}Qubits in use for each device (\texttt{ibm\_brisbane}, \texttt{ibm\_osaka} and \texttt{ibm\_cusco}) and their median key properties: single and two-qubit gate errors, readout errors, and T1 and T2 times. }
	\label{tab:device}
	\vspace{2mm}
	\begin{tabular}{lcccccc}
		\hline
		Backend & Qubits in use & 1q-gate err. & 2q-gate err. & Readout err. & T1[$\mu$s] &  T2[$\mu$s]\\
		\hline
		\texttt{brisbane} & \begin{tabular}{p{9cm}}[51, 36, 32, 31, 30, 17, 12, 11, 10, 9, 8, 7, 6, 5, 4, 3, 2, 1, 0, 14, 18, 19, 20, 21, 22, 23, 24, 25, 26, 27, 28, 35, 47, 48, 49, 55, 68, 69, 70, 74, 89, 88, 87, 86, 85, 73, 66, 65, 64, 54, 45, 44, 43, 42, 41, 40, 39, 38, 37, 52, 56, 57, 58, 71, 77, 76, 75, 90, 94, 95, 96, 109, 114, 115, 116, 117, 118, 119, 120, 121, 122, 123, 124, 125, 126, 112, 108, 107, 106, 105, 104, 103, 102, 92, 83, 82, 81, 80, 79, 78]\end{tabular}
    & 0.021\% & 0.690\% & 1.21\% & 251 & 175 \\
  		\texttt{osaka} & \begin{tabular}{p{9cm}}[2, 1, 0, 14, 18, 19, 20, 21, 22, 15, 4, 5, 6, 7, 8, 9, 10, 11, 12, 17, 30, 31, 32, 36, 51, 50, 49, 55, 68, 69, 70, 74, 89, 88, 87, 86, 85, 73, 66, 65, 64, 54, 45, 46, 47, 35, 28, 27, 26, 25, 24, 34, 43, 42, 41, 40, 39, 38, 37, 52, 56, 57, 58, 71, 77, 76, 75, 90, 94, 95, 96, 109, 114, 115, 116, 117, 118, 110, 100, 99, 98, 91, 79, 80, 81, 82, 83, 92, 102, 103, 104, 111, 122, 123, 124, 125, 126, 112, 108, 107]\end{tabular}
    & 0.029\% & 0.779\% & 2.46\% & 261 & 138 \\
		\texttt{cusco} & \begin{tabular}{p{9cm}}[13, 12, 11, 10, 9, 8, 7, 6, 5, 4, 3, 2, 1, 0, 14, 18, 19, 20, 21, 22, 23, 24, 34, 43, 42, 41, 40, 39, 38, 37, 52, 56, 57, 58, 59, 60, 61, 62, 72, 81, 82, 83, 84, 85, 73, 66, 65, 64, 54, 45, 46, 47, 35, 28, 29, 30, 31, 32, 36, 51, 50, 49, 55, 68, 69, 70, 74, 89, 88, 87, 93, 106, 107, 108, 112, 126, 125, 124, 123, 122, 121, 120, 119, 118, 117, 116, 115, 114, 109, 96, 95, 94, 90, 75, 76, 77, 78, 79, 91, 98]\end{tabular}
    & 0.040\% & 1.370\% & 3.46\% & 138 & 87 \\
		\hline
	\end{tabular}
\end{table*}

\end{document}